\documentclass[12pt]{article}
\usepackage[T1]{fontenc}
\usepackage[utf8]{inputenc}
\usepackage[italian,english]{babel}           
\usepackage{makeidx}
\usepackage{graphicx}
\usepackage{subfigure}
\usepackage{multicol} 
\usepackage{amsmath, bm, physics, amssymb, mathtools, gensymb}
\usepackage{nicefrac,xspace}
\usepackage[textsize=small,figwidth=\linewidth]{todonotes}
\usepackage[output-decimal-marker={.},product-units=single]{siunitx}
\usepackage[pdfborder={0 0 0.25}]{hyperref}       
\usepackage{newtxtext}       %
\usepackage{newtxmath}       
\makeindex             

\usepackage{geometry}
\newcommand{\Eq}[2][]{Eq.~\eqref{#2}\textsubscript{#1}}   
\newcommand{\Eqs}[1]{Eqs.~\eqref{#1}}                     
\newcommand{\Sec}[1]{Section~\ref{#1}}                     
\newcommand{\Fig}[2][]{Figure~\ref{#2}{#1}}                 
\newcommand*{\eg}{e.g.\@\xspace}                                     
\newcommand*{\ie}{i.e.\@\xspace}                                     
\newcommand{\PieD}{\boldsymbol{\Pi}^{\text{e}}_{\text{D}}}
\newcommand{\PieDxx}{\Pi^{\text{e}}_{\text{D}11}}
\newcommand{\PieDyy}{\Pi^{\text{e}}_{\text{D}22}}
\newcommand{\PieDzz}{\Pi^{\text{e}}_{\text{D}33}}
\newcommand{\PieDyz}{\Pi^{\text{e}}_{\text{D}23}}
\newcommand{\F}{\textbf{F}}
\newcommand{\B}{\textbf{B}}

\DeclareMathOperator{\dev}{dev}

\begin{document}

\title{Foundations of viscoelasticity and application to soft tissues mechanics}
\date{}
\author{Michele Righi and Valentina Balbi}
%
%
\maketitle

\abstract{Soft tissues are complex media, they display a wide range of mechanical properties such as anisotropy and non-linear stress-strain behaviour. They undergo large deformations and they exhibit a time-dependent mechanical behaviour, \ie they are viscoelastic. In this chapter we review the foundations of the linear viscoelastic theory and the theory of Quasi-Linear Viscoelasticity (QLV) in view of developing new methods to estimate the viscoelastic properties of soft tissues through model fitting. To this aim, we consider the simple torsion of a viscoelastic Mooney-Rivlin material in two different testing scenarios: step-strain and ramp tests. These tests are commonly performed to characterise the time-dependent properties of soft tissues and allow to investigate their stress relaxation behaviour. Moreover, commercial torsional rheometers measure both the torque and the normal force, giving access to two sets of data. We show that for a step test, the linear and the QLV models predict the same relaxation curves for the torque. However, when the strain history is in the form of a ramp function, the non-linear terms appearing in the QLV model affect the relaxation curve of the torque depending on the final strain level and on the rising time of the ramp. Furthermore, our results show that the relaxation curve of the normal force predicted by the QLV theory depends on the level of strain both for a step and a ramp tests. To quantify the effect of the non-linear terms, we evaluate the maximum and the equilibrium (as $t\!\rightarrow\!\infty$) values of the relaxation curves. Our results provide useful guidelines to accurately fit QLV models in view of estimating the viscoelastic properties of soft tissues.}

\section{Introduction}\label{sec:Intro}
Soft tissues, such as the brain, the skin, tendons and ligaments are viscoelastic materials, their mechanical behaviour is therefore time-dependent. Two typical experiments that show the time-dependent nature of soft tissues consists in stress relaxation and creep tests. In a stress relaxation test the tissue is suddenly stretched and then held in position for a certain time while the resulting stress is measured. Conversely, in a creep test, the load is applied to the tissue and the resulting deformation is measured. For many soft tissues the stress-relaxation curve has a decaying exponential form. Stress relaxation has been observed in the brain \cite{Budday2017,Rachid2014}, in ligaments and tendons \cite{DeVita2020,Shearer2020} and in the skin \cite{SkinTest2018}. At the microscale, the physical mechanisms behind stress relaxation differ from tissue to tissue. In tendons, for example, it has been observed that crimping and un-crimping of the hierarchical structures that build up the tissues, \ie the individual collagen fibrils, are responsible for the stress relaxation of the tissue \cite{TendonExp1959}. In the skin, the interaction between collagen and elastic fibers plays a crucial role in determining the time-dependent behaviour of the tissue. When the tissue is deformed, the cross-links maintain the structure and allow the elastic fibers to stretch and relax \cite{SkinVisco2013}.\\
However, in practice there is no machine that can instantaneously deform a tissue. A more realistic test is indeed a ramp test, where the tissue is deformed in a finite time and then held in that position. The duration of the ramp phase is called rising time $t^*$. When the rising time of the ramp is nearly zero, the ramp test can be well approximated by a step-strain test. However, if $t^*$ is not small (compared to the characteristic time constants of the material) modelling the ramp-test as a step-test can introduce errors in the estimation of the viscoelastic parameters.\\
From the modelling viewpoint, the simplest constitutive theory that can be used to describe the time-dependent behaviour of soft tissues is the linear viscoelastic theory, where the stress is related to the strain by a time-dependent function which in turn depends on the tissue's viscoelastic parameters. Linear models are based on three main assumptions:
\begin{enumerate}
	\item  the tissue remembers the past deformation history through a \textit{fading memory}, so that contributions to recent strain increments are more important than past contributions. A typical form of the time-dependent parameters that satisfies this assumption is a decaying exponential form;
	\item according to the Boltzmann superposition principle, the total stress at the current time $t$ is given by the sum of all past stress contribution;
	\item the deformation applied to the tissue is small.
\end{enumerate}
In early times, linear models have been employed to predict the viscoelastic behaviour of soft tissues. However, soon scientists have realised that these models do not provide accurate predictions, mainly because in reality soft tissues undergo large deformations. To overcome this limitation, Fung proposed what is now called the Quasi Linear Viscoelastic (QLV) theory, which is the simplest extension of the linear theory to large deformations \cite{Fung2013}. QLV models can capture stress-relaxation and creep, the strain-rate dependent response and account for large deformations. Moreover, the governing equations of the viscoelastic problem can be solved analytically for the most common modes of deformations used in experiments (\eg tension, compression, equi-biaxial tension, simple shear and torsion). Therefore, the constitutive parameters can be directly estimated through fitting of the experimental data, by implementing a minimisation algorithm. Linear and QLV models have a common limitation: being based on the linear superposition principle, they cannot account for the coupling between different time-scales, which is a limitation, especially for tissues with hierarchical structures. Although more complex non-linear models that account for this coupling have been proposed, they are numerically costly when it comes to model fitting and material parameters estimation \cite{Green1957,Findley2013}. Another class of non-linear models goes under the umbrella of internal variable or rate-type models which have recently gained popularity among the biomechanical community \cite{Reese2003,Reese1998,Lion1997}. These models are based on thermodynamics foundations. According to the multiplicative decomposition, the gradient of the deformation is split into an elastic and a viscous part. The resulting stress is then split into the sum of an elastic and a viscous term. The elastic stress is generally written with respect to an elastic strain energy function. The viscous stress is written with respect to a number of internal variables, whose evolution laws are dictated by the second law of thermodynamics and motivated by the linear theory \cite{Valanis2014}. This approach has the advantage of allowing an easy implementation of the constitutive model into finite element codes. However, when it comes to model fitting, the resulting equations are in implicit forms and need to be solved numerically, even for simple deformation modes.\\
Finally, differential-type models formulate the time-dependent constitutive equation in terms of the derivatives of the right stretch tensor evaluated at the current time \cite{pucci-diff}. Despite being computationally easy to implement, these models do not allow for an explicit form with respect to the relaxation functions, therefore they are less straightforward to fit with experimental data.
We conclude this brief introductory review by noting that viscoelasticity is not the only time-dependent property of soft tissues. Rate-type effects, such as stiffening and softening as a results of cyclic loading and unloading and ageing are other common effects displayed by biological tissues \cite{Afsar2020,Kazeroon2020}.\\
	In this chapter, we focus on viscoelasticity with the aim of providing useful guidelines on model fitting and estimation of the viscoelastic parameters for linear and QLV models. We consider two main experimental scenarios, the step-and-hold test and the ramp-and-hold test for the torsion of a cylindrical tissue. These tests are common experimental protocols used to investigate the viscoelastic properties of soft tissues, in particular stress-relaxation. In \Sec{sec:Linear} we review the standard linear viscoelastic theory and its rheological interpretation. In \Sec{sec:QLV} we review the QLV theory following the formulation proposed in \cite{DePascalis2014}. In \Sec{sec:torsion}, we consider the simple torsion of a cylindrical sample. This deformation can be performed with commercially available rheometers which measure both the torque and the normal force required to twist a cylindrical sample, giving access to two independent sets of data. Torsion has been successfully used to characterise the elastic properties of the brain in large deformations \cite{Balbi2019} suggesting that the tissue behaves as a Mooney-Rivlin material. The same behaviour was previously observed in simple shear experiments \cite{RachidShear2013}. In view of applications to brain mechanics, we therefore solve the equilibrium equations for a viscoelastic material whose elastic stress obeys a Mooney-Rivlin law and we calculate the expressions for the torque and the normal force. In \Sec{sec:results} we compare the predictions of the QLV model in the scenario of a step-strain test and of a ramp test. We conclude the chapter by discussing our results and by summarising the main findings.

\section{Linear viscoelastic models}\label{sec:Linear}
\begin{figure}[htb]
	\centering
	\includegraphics[width=0.55\linewidth]{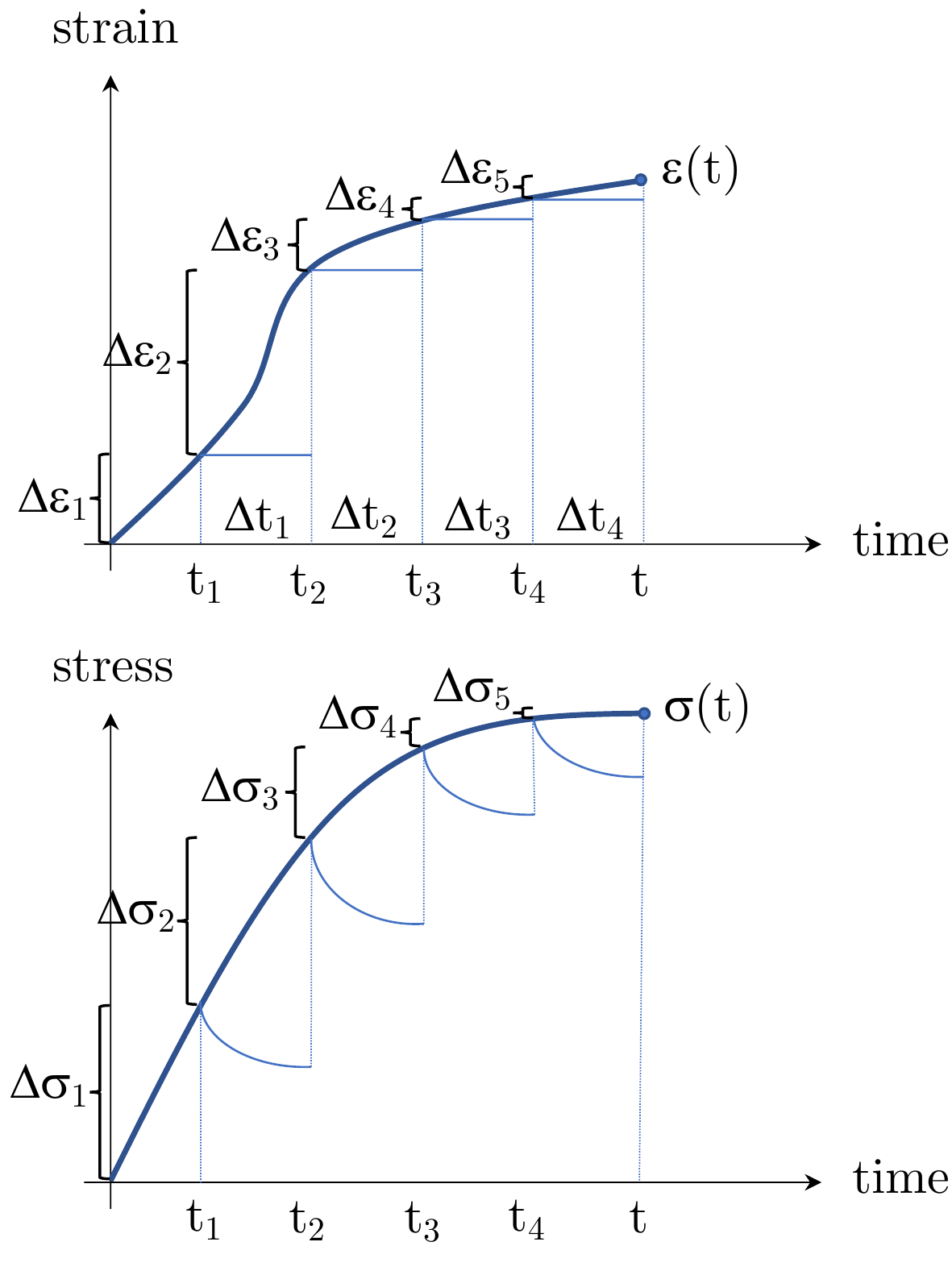}\caption{Boltzmann superposition principle: the strain history $\varepsilon$ is approximated as sum of steps $\Delta\varepsilon_i$ and the resulting total stress response $\sigma$ is the sum of the stress responses $\Delta\sigma_i$ to each step increment.}\label{fig:Boltz}
\end{figure}

Linear viscoelastic constitutive models are formulated by introducing the time dependency in the material parameters, a sort of fading memory which remembers the strain history of the material up to the current configuration. Accordingly, Hooke's law $\boldsymbol{\sigma}=\mathbb{K}:\bm{\varepsilon}$ rewrites as follows:
\begin{equation}\label{eq:lin-ve}
	\boldsymbol{\sigma}(t)=\int_{-\infty}^{t} \mathbb{K}(t-\tau): \dv{\bm{\varepsilon}(\tau)}{\tau} \dd{\tau},
\end{equation}
where $\boldsymbol{\sigma}$ is the stress tensor, $\boldsymbol{\varepsilon}$ is the infinitesimal strain tensor, and $\mathbb{K}(t)$ is called the \textit{tensorial relaxation function} and is a fourth-order tensor whose entries are the time-dependent material parameters. The symbol $:$ denotes the double contraction between a fourth-order tensor $\mathbb{Y}$ and a second-order tensor $\textbf{Z}$ such that $(\mathbb{Y}\!:\!\textbf{Z})_{ab}=Y_{abcd}Z_{cd}$.\\
\Eq{eq:lin-ve} is based on the Boltzmann superposition principle \cite{Boltz1874,Markovitz1977}. Accordingly, the total stress at the current time $t$ can be written as sum of past stress contributions up to the time $t$. In a one-dimensional setting, for the generic component $\sigma$ we can then write $\sigma=\sum_i\Delta\sigma_i$, as sketched in \Fig{fig:Boltz}.
Each $\Delta\sigma_i$ is the stress response to the step increment  $\Delta\varepsilon_i=\dv{\varepsilon}{t}\Delta t_i$ and is governed by the relaxation function $k_{\text{step}}(t-t_i)$.
Therefore, each stress increment can be written as $\Delta\sigma_i=k_{\text{step}}(t-t_i)\Delta\varepsilon_i$. By assuming that the strain history is continuous over time, the sum can be converted into an integral over time. The stress component $\sigma$ at time $t$ is then given by the following convolution integral:
\begin{equation}\label{eq:sigma1dof}
	\sigma(t)=\int_{-\infty}^{t} k_{\text{step}}(t-\tau) \dv{\varepsilon(\tau)}{\tau} \dd{\tau},
\end{equation}
for a given deformation history $\varepsilon(t)$. \\
\Eq{eq:lin-ve} is the tensorial version of \Eq{eq:sigma1dof}. The components of the tensor $\mathbb{K}(t)$ are the different relaxation functions of the tissue, \ie the time-dependent mechanical parameters.
We will show in \Sec{sec:bases} that the tensor $\mathbb{K}(t)$ can be split into its components according to a set of fourth-order bases. Thus, the resulting constitutive equation can be written with respect to different sets of relaxation functions according to the choice of bases.\\
From an experimental viewpoint, the choice of the bases might be dictated by which mechanical properties we want to estimate. For instance, if we are interested in estimating the time-dependent shear modulus we will perform a simple shear test or a torsion test, whereas if we want to estimate the time-dependent Young's modulus we will perform a tensile test. Moreover, in order to be able to estimate the components of $\mathbb{K}(t)$ we first have to specify their functional form with respect to time. In \Sec{sec:rel-fun} we will review a common form used in the biomechanics community, \ie the \text{Prony series} form. Furthermore, we will discuss two common experimental protocols that are performed to estimate the components of $\mathbb{K}(t)$, namely the step-strain and the ramp tests.

\subsection{Bases decomposition for the tensor $\mathbb{K}(t)$}\label{sec:bases}
In this section, we focus on the tensorial nature of the relaxation function $\mathbb{K}(t)$ and we show that the constitutive equation \eqref{eq:lin-ve} can be written with respect to different sets of components of $\mathbb{K}(t)$, i.e. the time-dependent mechanical properties of the tissue, according to different choices of fourth-order tensorial bases.
To simplify the analysis, in this chapter we restrict out attention to homogeneous isotropic tissues. The mechanical behaviour of such tissues is fully described by two independent mechanical parameters, \eg the bulk and the shear modulus, $\kappa(t)$ and $\mu(t)$ respectively, or the first Lam\'e parameter $\lambda(t)$ and the shear modulus. According to the elasticity theory, the elasticity tensor $\mathbb{C}$ for a homogeneous isotropic material depends only on two independent elastic constants. Similarly, the tensorial relaxation function $\mathbb{K}(t)$ for an isotropic material has two independent components $K_1(t)$ and $K_2(t)$ with respect to two bases $\mathbb{I}_1$ and $\mathbb{I}_2$, respectively, such that:
\begin{equation}\label{eq:gen-K}
	\mathbb{K}(t)=\sum_{n=1,2}K_n(t)\mathbb{I}_n,
\end{equation}
A well-known form of the constitutive equation for a homogeneous isotropic and compressible material follows by splitting the infinitesimal strain tensor $\bm{\varepsilon}$ into its hydrostatic and deviatoric parts. The hydrostatic part is associated to volume changes, whereas the deviatoric part is associated to the volume-preserving part of the deformation. The following set of bases splits the strain tensor into its hydrostatic and deviatoric parts:
\begin{equation}\label{eq:bases1}
	\mathbb{I}_{1abcd}=\dfrac{1}{3}\delta_{ab}\delta_{cd}\qquad \text{and}\qquad \mathbb{I}_{2abcd}=\dfrac{1}{2}\left(\delta_{ac}\delta_{bd}+\delta_{ad}\delta_{bc}\right)-\dfrac{1}{3}\delta_{ab}\delta_{cd}.
\end{equation}
Accordingly, \Eq{eq:lin-ve} takes the following form:
\begin{equation}\label{eq:VEbases1}
	\begin{split}
		\boldsymbol{\sigma}(t)&=\int_{-\infty}^{t} \sum_{n=1,2}K_n(t-\tau)\mathbb{I}_n: \dv{\boldsymbol{\varepsilon}(\tau)}{\tau} \dd{\tau}\\
		&=\int_{-\infty}^{t} K_1(t-\tau) \dv{\mathbb{I}_1:\boldsymbol{\varepsilon}(\tau)}{\tau} \dd{\tau}+\int_{-\infty}^{t} K_2(t-\tau) \dv{\mathbb{I}_2:\boldsymbol{\varepsilon}(\tau)}{\tau} \dd{\tau}\\
		&=\int_{-\infty}^{t} K_1(t-\tau) \dv{}{\tau}\left(\dfrac{1}{3}\tr\left( \boldsymbol{\varepsilon}(\tau)\right) \textbf{I}\right) \dd{\tau}+\int_{-\infty}^{t} K_2(t-\tau) \dv{}{\tau}\left(\boldsymbol{\varepsilon}(\tau)-\dfrac{1}{3}\tr\left( \boldsymbol{\varepsilon}(\tau)\right) \textbf{I}\right) \dd{\tau}\\
		&=\int_{-\infty}^{t} \kappa(t-\tau) \dv{}{\tau}\left(\tr\left( \boldsymbol{\varepsilon}(\tau)\right) \textbf{I}\right) \dd{\tau}+2\int_{-\infty}^{t} \mu(t-\tau) \dv{}{\tau}\left(\dev\left( \boldsymbol{\varepsilon}(\tau)\right) \right) \dd{\tau},
	\end{split}
\end{equation}
where $\delta_{ab}$ is the Kronecker delta ($\delta_{ab}\!=\!1$ if $a\!=\!b$ and $\delta_{ab}\!=\!0$ if $a\!\neq\!b$), $\textbf{I}$ is the second-order identity tensor and $\dev\boldsymbol{\varepsilon}=\boldsymbol{\varepsilon}-\frac{1}{3}\tr\left( \boldsymbol{\varepsilon}\right) \textbf{I}$ is the deviatoric part of the second-order tensor $\bm{\varepsilon}$. The bases $\mathbb{I}_1$ and $\mathbb{I}_2$ defined in \Eq{eq:VEbases1} act on a second-order tensor by splitting the tensor into its spherical and deviatoric parts, respectively. The associated material parameters $\kappa(t)$ and $\mu(t)$ are the time-dependent bulk and shear modulus, respectively.\\
For incompressible materials, \ie materials that deform by keeping their volume constant, the bulk modulus is much greater than the shear modulus ($\kappa(t)\gg \mu(t)$ for $\forall t$). Moreover, the following assumptions are true: $\tr\bm{\varepsilon}(t)\rightarrow 0$, $\kappa(t)\rightarrow \infty, \forall t$. In these limits, \Eq{eq:VEbases1} reduces to:
\begin{equation}\label{eq:linconst}
	\boldsymbol{\sigma}=-p(t)\textbf{I}+2\int_{-\infty}^{t} \mu(t-\tau) \dv{}{\tau}\left(\dev \boldsymbol{\varepsilon}(\tau) \right) \dd{\tau},
\end{equation}
where we have introduced the Lagrange multiplier $p(t)$:
\begin{equation}\label{eq:p}
	-p(t)=\lim_{\tr\bm{\varepsilon}(t)\rightarrow 0} \lim_{\kappa(t)\rightarrow \infty}
	\int_{-\infty}^{t} \kappa(t-\tau) \dv{}{\tau}\left(\tr \boldsymbol{\varepsilon}(\tau)\right) \dd{\tau}.
\end{equation}
The scalar $p(t)$ can be interpreted as a hydrostatic pressure and can be calculated by solving the governing equations of motions for a continuum body, upon imposing the boundary conditions.
Note that the stress component $-p(t) \textbf{I}$ represents a workless reaction with respect to the kinematic constraint of the deformation field. No dissipation is involved in the isochoric deformation of the body. Hence, for materials that can be treated as incompressible, only the deviatoric part of the stress exhibits a viscoelastic nature.

Similarly, by choosing the following bases
\begin{equation}
	\mathbb{J}_{1abcd}=\delta_{ab}\delta_{cd} \qquad \text{and} \qquad \mathbb{J}_{2abcd}=\dfrac{1}{2}\left(\delta_{ac}\delta_{bd}+\delta_{ad}\delta_{bc}\right),
\end{equation}
$\mathbb{K}(t)=\sum_{n=1,2}A_n(t)\mathbb{J}_n$ and \Eq{eq:lin-ve} writes as follows:
\begin{equation}\label{eq:VEbases2}
	\begin{split}
		\boldsymbol{\sigma}(t)&=\int_{-\infty}^{t} \sum_{n=1,2}A_n(t-\tau)\mathbb{J}_n: \dv{\boldsymbol{\varepsilon}(\tau)}{\tau} \dd{\tau}=\cdots=\\
		&=\int_{-\infty}^{t} A_1(t-\tau) \dv{}{\tau}\left(\tr\boldsymbol{\varepsilon}(\tau)\textbf{I}\right) \dd{\tau}+\int_{-\infty}^{t} A_2(t-\tau) \dv{}{\tau}\left(\boldsymbol{\varepsilon}(\tau)\right) \dd{\tau}\\
		&=\int_{-\infty}^{t} \lambda(t-\tau) \dv{}{\tau}\left(\tr\boldsymbol{\varepsilon}(\tau)\textbf{I}\right) \dd{\tau}+2\int_{-\infty}^{t} \mu(t-\tau) \dv{}{\tau}\left(\boldsymbol{\varepsilon}(\tau)\right) \dd{\tau}.
	\end{split}
\end{equation}
Now, we have $A_1(t)=\lambda(t)$, which is the time-dependent first Lam\'e parameter and $A_2(t)=2\mu(t)$. Moreover, note that $\mathbb{J}_1=3\mathbb{I}_1$ and $\mathbb{J}_2=\mathbb{I}_2-\mathbb{I}_1$ and the following link is true $\lambda(t)=\kappa(t)-\frac{2}{3}\mu(t)$, $\forall t$. Clearly, \Eqs{eq:VEbases1} and \eqref{eq:VEbases2} are equivalent forms of the constitutive equation \eqref{eq:lin-ve} and predict the same stress response to a general strain input $\bm{\varepsilon}(t)$. The choice of the bases and therefore the final form of the constitutive model is usually dictated by what type of material properties we want to determine and the type of experimental devices available for testing (\eg tensile machines, rheometers, bi-axial devices, etc.). In \Sec{sec:QLV} we will use the bases decomposition of \Eq{eq:bases1} to write the constitutive equation for the QLV model. In the next section, we focus on the mathematical form of the components of the tensor $\mathbb{K}(t)$.

\subsection{Rheological models for the relaxation function}\label{sec:rel-fun}
In order to use the constitutive equation \eqref{eq:lin-ve} for model fitting and parameter estimations, a mathematical form for the components of the tensorial function $\mathbb{K}(t)$ has to be chosen. On the one hand, the form of the relaxation functions is restricted by the following physical principles: positive strain energy and satisfaction of the second law of thermodynamics. Imposing the energy density to be non-negative during the tissue relaxation results in requiring the relaxation function to be positive $\forall\, t$ \cite{Lakes2009}. Furthermore, the second law of thermodynamics (dissipation inequality) requires that the relaxation function decreases monotonically with time \cite{Lakes2009}. On the other hand, any function that satisfies the physical constraints and replicates the shape of the observed stress relaxation curve can be used. From the experimental viewpoint, a classical experiment that can be done to determine the relaxation curve is to apply a displacement to the tissue, then hold the tissue in position (\ie maintain a constant level of strain) for a certain time and measure the resulting stress curve. For most soft tissues, the measured stress relaxation curve has a decaying exponential behaviour \cite{Chatelin2010,DeVita2020,Shearer2020}.

\begin{figure}[htb]
	\centering
	\includegraphics[width=0.55\linewidth]{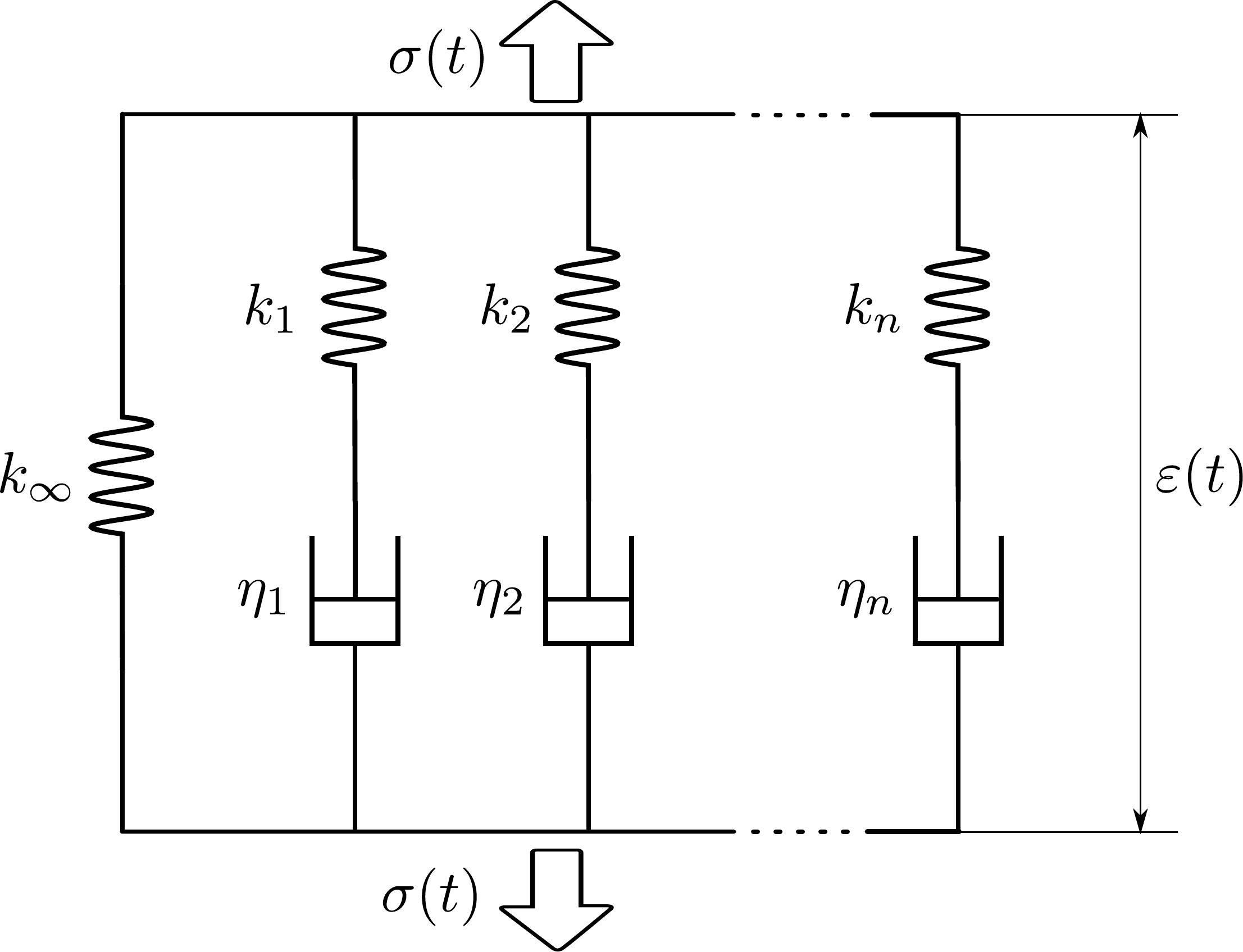}
	\caption{Rheological models: springs and dash-pots arrangement of the generalized Maxwell model. The parameters $k_i$ are spring constants (\si{Pa}) and $\eta_i$ are dash-pot constants (\si{\Pa \second}), or viscosities. $\sigma$ is the stress response of the system to the applied strain $\varepsilon$.}
	\label{fig:gen-max}
\end{figure}

The simplest form for the relaxation function that captures the exponential decaying behaviour and satisfies the physical constraints is the so-called Prony series. The Prony-series has its origin in one-dimensional rheological models \cite{Lakes2009,Fung2013}. Such models are represented by an arrangement of linear springs and linear dash-pots. The layout of such arrangements of elements provides the qualitative behaviour of the system, \eg solid-like or fluid-like behaviour, while the values of the constants characterise the quantitative behaviour. In \Fig{fig:gen-max} we sketch the generalized Maxwell scheme, which is used to model the viscoelastic response of solid materials. The isolated spring $k_{\infty}$ represents the residual (long-term) elasticity of the tissue.

One-dimensional rheological models can be described by a linear ordinary differential equation in the variables $\sigma$ (the stress) and $\varepsilon$ (the strain).
The stress response $\sigma$ of the system to a step-strain input $\varepsilon$ provides the form of the relaxation function $k_{\text{step}}$ in \eqref{eq:sigma1dof}.
A convenient way to derive the response of rheological models, especially when a large number of elements is involved, is to employ the Laplace transform.
The approach involves the following steps:
\begin{enumerate}
	\item Write the constitutive equations for all the elements in the system: $\sigma_i=k_i\varepsilon_i$ for springs and $\sigma_i=\eta_i\dot{\varepsilon}_i$ for dash-pots. Then write the equilibrium equations $\sigma=\sum_i\sigma_i$ for elements in parallel that experience the same strain and $\varepsilon=\sum_i\varepsilon_i$ for elements in series that experience the same stress.

	\item By applying the Laplace transform to each equations at point 1, convert the system of mixed differential and algebraic equations into a system of only algebraic equations.

	\item Apply a variable elimination procedure to the system derived at point 2. The system has $3n+2$ equations in $3n+3$ unknowns, \ie $\varepsilon_i$ and $\sigma_i$. This reduces the system to the single equation: $\bar{\sigma}(s)=F(s)\bar{\varepsilon}(s)$, where $\bar{\sigma}$ and $\bar{\varepsilon}$ are the Laplace transforms of $\sigma(t)$ and $\varepsilon(t)$, respectively. $F(s)$ is the transfer function in the complex domain represented by the complex variable $s$.

	\item By applying the inverse Laplace transform to the equation $\bar{\sigma}(s)=F(s)\bar{\varepsilon}(s)$, obtain the constitutive equation in the time domain.
\end{enumerate}

From point 3, the transfer function $F(s)$ of the generalized Maxwell model in \Fig{fig:gen-max} is given by:
\begin{equation}\label{eq:TF-gen-max}
	F(s)=\frac{\bar{\sigma}(s)}{\bar{\varepsilon}(s)}=k_{\infty}+\sum_{i=1}^{n}\frac{s \eta_i}{1+s\tau_i},
\end{equation}
where the constants $\tau_i=\frac{\eta_i}{k_i}$ are called relaxation times of the model.

To calculate the response of a generalised Maxwell system to a step input, we write the strain $\varepsilon(t)=\varepsilon_0H(t)$, where $H(t)$ is the Heaviside function: $H(t)=1$, $\forall t\geq0$ and $H(t)=0$, $\forall t<0$ and $\varepsilon_0$ is the amplitude of the step. Then we calculate the Laplace transform $\bar{\varepsilon}$ and substitute the result into \Eq{eq:TF-gen-max} obtaining:
\begin{equation}\label{eq:sigmaC}
	\bar{\sigma}(s)= \left(\frac{k_{\infty}}{s}+\sum_{i=1}^{n}\frac{\eta_i}{1+s\tau_i}\right)\varepsilon_0.
\end{equation}
According to point 4, by transforming back into the time domain, we obtain the stress response to a step-strain input with amplitude $\varepsilon_0$:
\begin{equation}\label{eq:stepStress}
	\sigma(t)=\left( k_{\infty}+\sum_{i=1}^{n}k_i e^{-\frac{t}{\tau_i}}\right)\varepsilon_0.
\end{equation}
From \Eq{eq:stepStress} we can then calculate the stress response to a step-strain input, i.e. the relaxation function $k_{\text{step}}(t)$, by dividing $\sigma(t)$ by the amplitude of the step, as follows:
\begin{equation}\label{eq:prony}
	k_{\text{step}}(t)=\frac{\sigma(t)}{\varepsilon_0}=k_{\infty}+\sum_{i=1}^{n}k_i e^{-\frac{t}{\tau_i}}.
\end{equation}
The function \eqref{eq:prony} is called Prony series and is a sum of exponential terms, each corresponding to a branch of the generalised Maxwell in \Fig{fig:gen-max}. Note that the relaxation function $k_{\text{step}}(t)$ does not depend on the strain $\varepsilon$.

\begin{figure}[htb]
	\centering
	\includegraphics[width=0.55\linewidth]{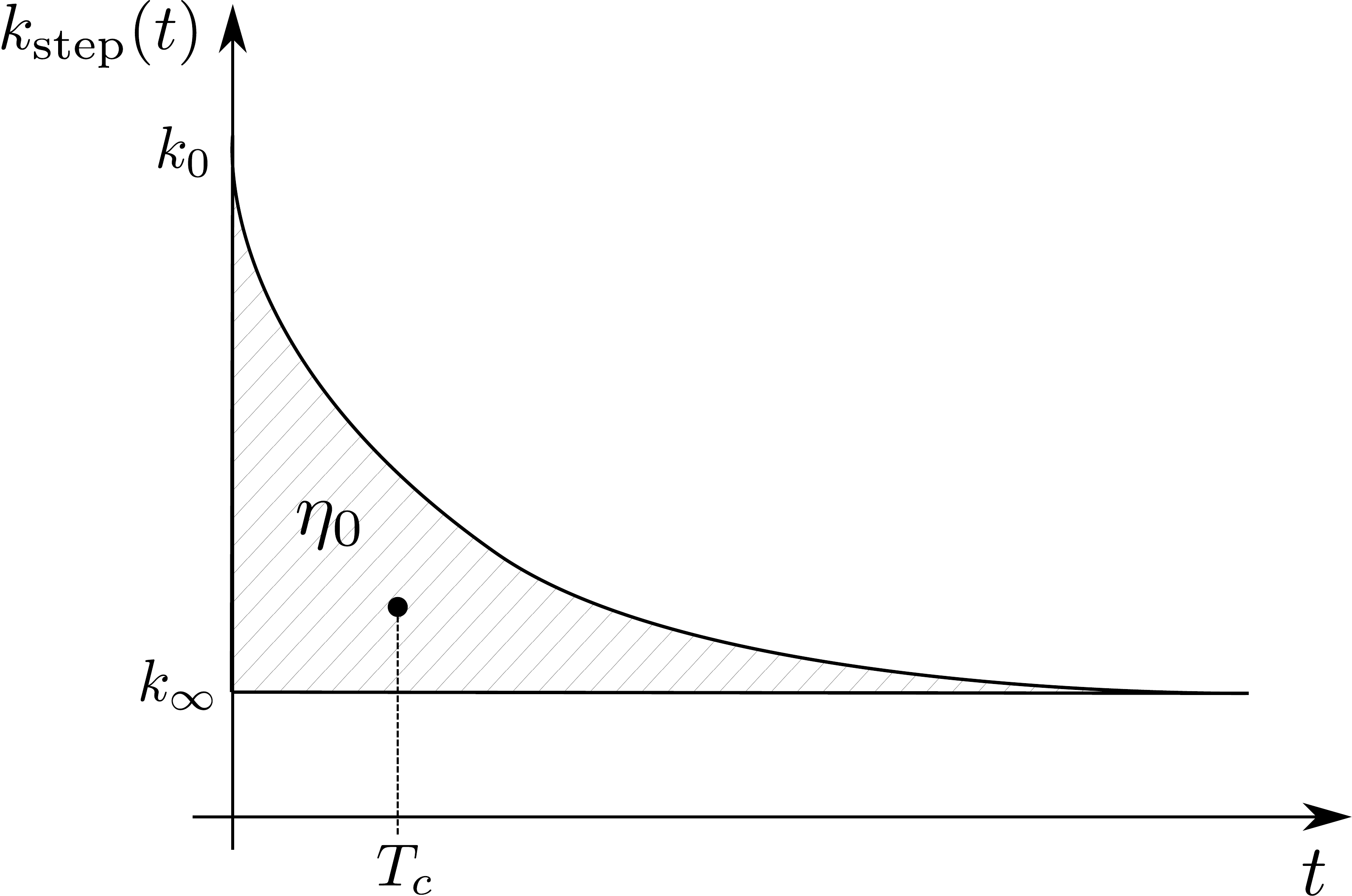}
	\caption{Stress response of the generalised Maxwell model to a step-strain input.}
	\label{fig:rel-fun}
\end{figure}

The constants $k_{\infty},k_i$ and $\eta_i$ can be determined by fitting \Eq{eq:prony} to the stress relaxation curve experimentally measured from a step-strain test. The sample is suddenly deformed up to the strain $\varepsilon_0$ and held in position for a certain amount of time.
In the limit $t\rightarrow\infty$ \Eq{eq:stepStress} recovers the elastic equilibrium stress ($\sigma_{\infty}\!=\!k_{\infty}\varepsilon_0$) and the relaxation function in \Eq{eq:prony} reduces to the long term elastic modulus $k_{\infty}$:
\begin{equation}\label{keyinf}
	k_{\infty}=\lim_{t\rightarrow\infty} k_{\text{step}}(t).
\end{equation}
Experimentally, this limit is equivalent to a very slow ramp test, \ie a quasi-static test, where the final value of strain $\varepsilon_0$ is attained as $t\!\rightarrow\!\infty$.\\
On the other hand, in the limit $t\rightarrow0$,  \Eq{eq:prony} reduces to the instantaneous elastic modulus $k_0$:
\begin{equation}\label{eq:k0}
	k_0=k_{\text{step}}(0)=k_{\infty}+\sum_{i=1}^n k_i,
\end{equation}
The value in \Eq{eq:k0} corresponds to the maximum of the relaxation function, see \Fig{fig:rel-fun}. Experimentally, this limit corresponds to the application of an instantaneous strain, which is practically impossible to perform.
The constants $k_0$ and $k_{\infty}$ are the elastic parameters of the constitutive model and describe the instantaneous and long-term elastic behaviours of the tissue, respectively. \\
The viscous behaviour of the tissue is associated to the parameters $\eta_i$ and $\tau_i$. To get some insights on the viscous parameters, we define the function:
\begin{equation}\label{ktilde}
	\tilde{k}_{\text{step}}(t)=k_{\text{step}}(t)-k_{\infty}=\sum_{i=1}^{n}k_i e^{-\frac{t}{\tau_i}}
\end{equation} and we integrate over the whole time spectrum:
\begin{equation}\label{eq:etaSum}
	\eta_0=\int_{0}^{\infty}  \tilde{k}_{\text{step}}(t) \dd{t}=\sum_{i=1}^{n}\eta_i.
\end{equation}
The value $\eta_0$ has an important geometrical interpretation since it represents the area between the curve $k_{\text{step}}(t)$ and the asymptotic line $k_{\infty}$, see \Fig{fig:rel-fun}. The bigger the area, the more viscous the material.\\
Moreover, we can compute the mean relaxation time $T_c$ of the tissue as follows:
\begin{equation}\label{eq:Tc}
	T_c=\frac{\int_{0}^{\infty} t \tilde{k}_{\text{step}}(t) \dd{t}}{\int_{0}^{\infty}  \tilde{k}_{\text{step}}(t) \dd{t}}
	=\frac{\sum_{i=1}^{n}\eta_i \tau_i}{\sum_{i=1}^{n}\eta_i}.
\end{equation}
Geometrically, $T_c$ represents the centroid of the shaded area below the relaxation function $k_{\text{step}}(t)$ and the asymptotic value $k_{\infty}$, and can be interpreted as the average relaxation time.

\subsubsection*{Ramp tests}\label{sec:ramp}
\begin{figure}[htb]
	\centering
	\includegraphics[width=0.4\linewidth]{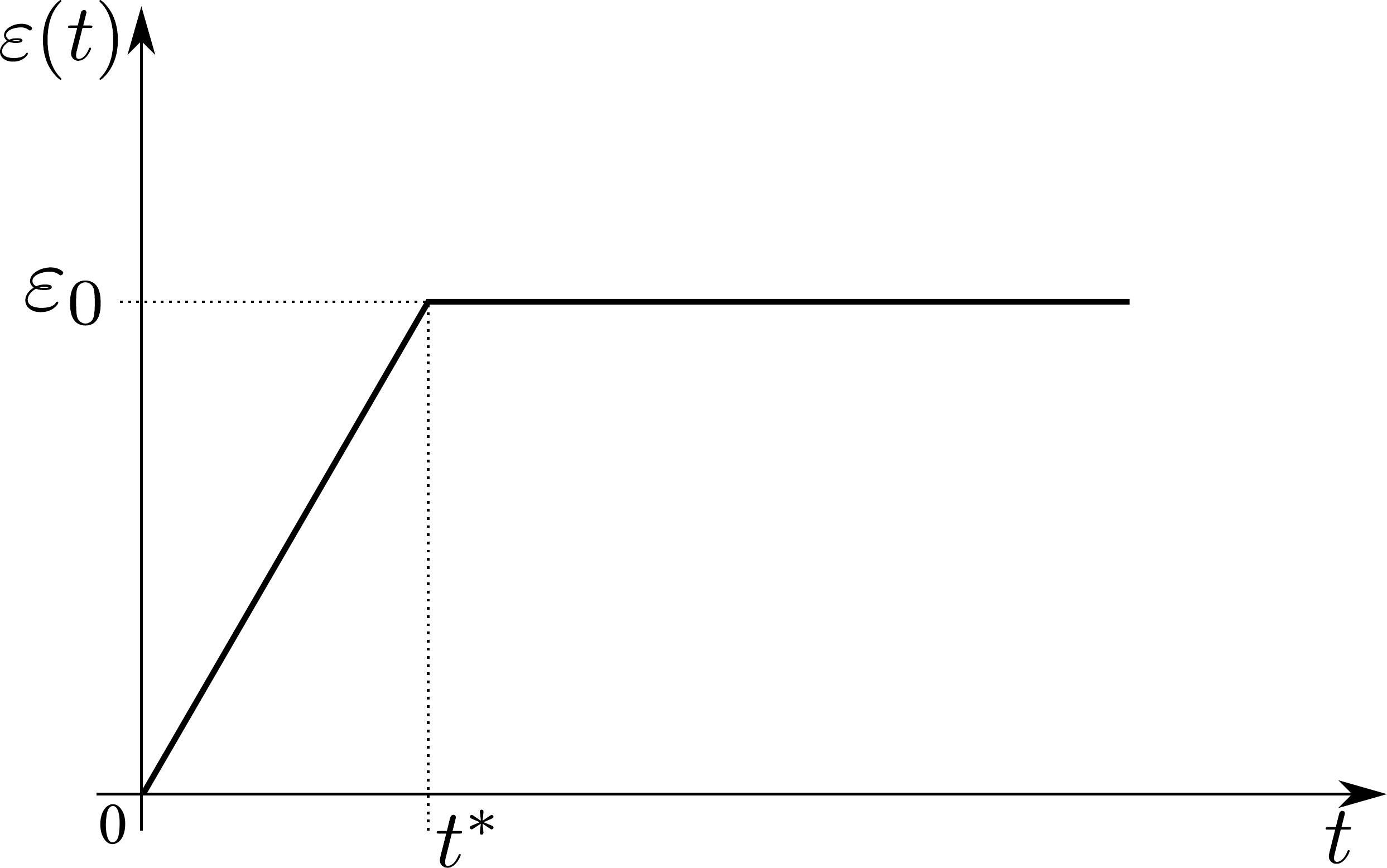}
	\caption{Strain history for a ramp and hold experiment. The constant strain value $\varepsilon_0$ is reached at the end of the loading phase ($t=t^*$), where the strain increases at a constant rate.}
	\label{fig:ramp}
\end{figure}

Now, we recall that Eqs.~(\ref{eq:prony}-\ref{eq:Tc}) are valid for a strain input in the form of a step function (\ie $\varepsilon(t)=H(t)\varepsilon_0$), where the strain value $\varepsilon_0$ is attained instantaneously. However, such experiment is not feasible in laboratory since that would require a testing machine able to reach an infinite rate of deformation. Real tests are much closer to a ramp test, where the strain $\varepsilon_0$ is reached after a finite rising time $t^*>0$, as shown in \Fig{fig:ramp}.

In view of providing an analytical expression for the relaxation curve to be fitted with the experimental data, in this section we derive the stress response of a generalised Maxwell system to a ramp input.
The strain input for a ramp test takes the following form:
\begin{equation}\label{eq:ramp}
	\varepsilon(t)=\frac{\varepsilon_0}{t^*}t-\frac{\varepsilon_0}{t^*}(t-t^*)H(t-t^*),
\end{equation}
where $H(t)$ is the Heaviside function.
We can calculate the stress response to the input in \Eq{eq:ramp} by substituting \Eq{eq:prony} and \Eq{eq:ramp} into \Eq{eq:sigma1dof}. We then obtain:
\begin{equation}\label{eq:sigmaramp}
	\sigma(t)=k_{\text{ramp}}(t)\varepsilon_0, \qquad \text{for} \quad t> t^*,
\end{equation}
where:
\begin{equation}\label{eq:relaxramp}
	k_{\text{ramp}}(t)=\left(k_{\infty}+\frac{1}{t^*}\sum_{i=1}^{n}\tau_i k_i e^{-\frac{t}{\tau_i}}\left( e^{\frac{t^*}{\tau_i}} -1 \right)\right), \qquad \text{for} \quad t> t^*.
\end{equation}
Note that by taking the limit for $t\rightarrow\infty$ in \Eq{eq:relaxramp}, we recover the long term elastic modulus $k_{\infty}$. Similarly, the equilibrium stress as $t\rightarrow\infty$ is given by $\sigma_{\infty}\!=\!k_{\infty}\varepsilon_0$ as for the step test. \\
On the other hand, the stress response at $t\!=\!t^*$ is now affected by the previous deformation history at $t\!<\!t^*$.
To quantify the effect of the rising time of the ramp on the instantaneous elastic and viscous response, we calculate the function $\tilde{k}_{\text{ramp}}(t)$:
\begin{equation}\label{eq:kramp}
	\tilde{k}_{\text{ramp}}(t)=k_{\text{ramp}}(t)-k_{\infty}=\sum_{i=1}^{n}\frac{k_i}{\nu_i} e^{-\frac{t}{\tau_i}}\left( e^{\nu_i} -1 \right),
\end{equation}
where $\nu_i=t^*/\tau_i$ are the ratio between the rise time and the characteristic time constants of the tissue $\tau_i$. Note that by taking the limit $t^*\rightarrow0$ (\ie a step input), \Eq{eq:kramp} recovers \Eq{ktilde}:
\begin{equation}\label{key}
	\lim_{t^*\rightarrow 0} \tilde{k}_{\text{ramp}}(t)=\tilde{k}_{\text{step}}(t)=\sum_{i=1}^{n}k_i e^{-\frac{t}{\tau_i}}.
\end{equation}
Moreover, we define the modified instantaneous elastic modulus $k_{0\text{ramp}}$ as the function \eqref{eq:kramp} evaluated at the end of the ramp phase, \ie at $t=t^*$:
\begin{equation}\label{eq:k0ramp}
	k_{0\text{ramp}}=\tilde{k}_{\text{ramp}}(t^*)=\sum_{i=1}^{n}k_i \zeta_i=\bm{k} \cdot \bm{\zeta}, \quad \text{with} \quad
	\zeta_i=\nu_i^{-1}\left(1-e^{-\nu_i} \right).
\end{equation}
$\bm{k}$ and $\bm{\zeta}$ are vectors with components $k_i$ and $\zeta_i$, $i\!=\!\{1,\dots,n\}$, respectively.
Since $0\leq\nu_i<\infty$, then the parameters $\zeta_i$ range between $0<\zeta_i\leq1$. When $\nu_i=0$, $\zeta_i=1$ and the ramp recovers the perfect step-strain input and the elastic modulus in \Eq{eq:k0ramp} reduces to the instantaneous modulus in \Eq{eq:k0}.

We note that the elastic constants $k_i$ are intrinsic properties of the tissue, therefore they do not depend on the testing procedure nor on the form of the strain history or on the strain-rate at which the test is performed.
On the other hand it is well-known that the response of a viscoelastic material strongly depends on the strain-rate.
The coefficients $\zeta_i$ account for the strain rate of the deformation process, \ie for the fact that the strain is applied to the tissue in a finite time $t^*$. The vector $\bm{\zeta}$ allows to isolate the effect of the strain-rate from the constant elastic moduli $\bm{k}$ that describe the material. Therefore, fitting stress relaxation data by using the relaxation function of a step instead of the ramp will result in underestimating the elastic moduli.
The smaller the strain-rate (\ie the greater $t^*$), the smaller $k_{0\text{ramp}}$. In the limit of a quasi-static deformation, only the infinite modulus $k_{\infty}$ is recoverable.

Similarly to \Eq{eq:etaSum} we can now compute the total viscosity of the material using the modified function $\tilde{k}_{\text{ramp}}(t)$ in \Eq{eq:kramp}:
\begin{equation}\label{eq:etaSum0}
	\eta_{0\text{ramp}}=\int_{t^*}^{\infty}  \tilde{k}_{\text{ramp}}(t) \dd{t}=\sum_{i=1}^{n}\eta_i \zeta_i=\bm{\eta}\cdot\bm{\zeta}.
\end{equation}
\Eq{eq:etaSum0} highlights that the same set of coefficients $\zeta_i$ that link the elastic moduli also link the viscous constants.
Since for cases of practical interest $0<\zeta_i<1$, it follows that $\eta_{0\text{ramp}}<\eta_0$. Therefore, neglecting the influence of the deformation rate will result in underestimating the total viscosity of the material.
This is in agreement with the result $\tilde{k}_{0\text{ramp}}<\tilde{k}_0$, with $\tilde{k}_0=k_0-k_{\infty}$.\\
In conclusion, by fitting a stress relaxation curve obtained from a ramp test with \Eq{eq:prony} we can obtain a correct estimation of the infinite modulus $k_{\infty}$, which is strain-rate independent. However, the peak of the relaxation curve, which is related to the instantaneous response of the tissue and therefore to its instantaneous elastic modulus $k_0$, depends on the strain rate and it is given by \Eq{eq:k0ramp}. In particular, the lower the strain rate, the lower the peak. The area enclosed between the relaxation function and its horizontal asymptote is related to the total viscosity of the material. Since $k_{\infty}$ is not affected by the rate of deformation, $k_{0\text{ramp}}$ decreases with the area represented by $\eta_{0\text{ramp}}$.\\
These preliminary synthetic information ($k_{\infty}$, $k_{0\text{ramp}}$, $\eta_{0\text{ramp}}$) derived from observation of the experimental relaxation function can be used as a starting point of the fitting procedure to determine the constitutive parameters of the rheological model ($k_{\infty}$, $k_i$, $\tau_i$).

\section{QLV model}\label{sec:QLV}
The linear model in \Eq{eq:lin-ve} predicts accurate results only in the small deformation regime, \ie when $\varepsilon\approx0$. However it fails to accurately predict the stress response when a tissue is subjected to a large deformation. To account for large deformations, Fung originally proposed the theory of Quasi Linear Viscoelasticity (QLV) \cite{Fung2013}, which is the extension of the linear theory we reviewed in the previous section to the large deformation regime. In this section we review the QLV theory, following \cite{DePascalis2014} to derive the constitutive equation for isotropic compressible and incompressible soft tissues.\\
The QLV theory is based on the same assumptions of the linear theory, i.e. the Boltzmann superposition principle and the assumption of fading memory. Moreover, Fung postulated that the total stress is separable into the product of a function of time, \ie the relaxation function, and a function of the deformation, \ie the elastic stress. The former accounts for the time-decaying relaxation of the stress and the latter accounts for the non-linear elastic response of the tissue. In the QLV formulation, the relation between the elastic stress and the strain is non-linear.
To write the constitutive equation for a QLV model, we start by rewriting the linear model in \Eq{eq:lin-ve} in the following equivalent form:
\begin{equation}\label{eq:lin-sigma}
	\bm{\sigma}(t)=\int_{0}^t\mathbb{G}(t-\tau):\dfrac{\dd \bm{\sigma}^{\text{e}}(\tau)}{\dd \tau}\dd{ \tau}
\end{equation}
where the tensor $\mathbb{G}(t)$ is now a fourth-order tensor whose components are non-dimensional and such that $G_n(0)=1$ for $n=\{1,2\}$. We call $\mathbb{G}(t)$ the reduced relaxation tensor. Note that in \Eq{eq:lin-sigma} we have assumed that the deformation history starts at $t=0$. The stress term $\bm{\sigma}^{\text{e}}=\mathbb{K}(0)\bm{\varepsilon}$ is the linear elastic stress.
Fung proposed to replace the linear stress $\bm{\sigma}^{\text{e}}$ by the corresponding instantaneous elastic stress in large deformation and rewrite the constitutive equation \eqref{eq:lin-sigma} as follows:
\begin{equation}\label{QLV-initial}
	\bm{\Pi}(t)=\int_0^t\mathbb{G}(t-\tau):\dfrac{\dd \bm{\Pi}^{\text{e}}(\tau)}{\dd \tau}\dd{ \tau}.
\end{equation}
The tensor $\bm{\Pi}(t)$ is the second Piola-Kirchhoff stress tensor and $\bm{\Pi}^{\text{e}}(t)$ is the elastic second Piola-Kirchhoff stress tensor defined as follows:
\begin{equation}\label{Piola-elastic}
	\bm{\Pi}^{\text{e}}=J \textbf{F}^{-1}\textbf{T}^{\text{e}}\textbf{F}^{-\text{T}}.
\end{equation}
$\textbf{T}^{\text{e}}$ is the elastic Cauchy stress, $\textbf{F}=\textbf{I}-\nabla\textbf{u}=\partial \textbf{x}/\partial\textbf{X}$ is the deformation gradient associated to the large deformation $\textbf{x}=\chi(\textbf{X})$, and $J=\det\textbf{F}$. $\textbf{x}$ and $\textbf{X}$ are the position vectors in the undeformed and deformed configurations, respectively. We use the notation $\textbf{T}$ to avoid confusion with the linear stress tensor $\bm{\sigma}$. In the small deformation regime the undeformed and deformed configurations coincide since $\nabla\textbf{u}\approx0$ and $\textbf{F}\approx\textbf{I}$ and therefore the second Piola-Kirchhoff and the Cauchy stress tensors also reduce to the same stress tensor.\\
Now, we can use the bases in \Eq{eq:VEbases2} to split the tensor $\mathbb{G}$ of \Eq{QLV-initial}. We call the associated components $\mathcal{H}(t)$ and $\mathcal{D}(t)$, respectively, and we rewrite \Eq{QLV-initial} in the following form:
\begin{equation}\label{eq:QLV-final}
	\begin{split}
		\bm{\Pi}(t)&=\int_0^t \mathcal{H}(t-\tau)\dfrac{\dd \bm{\Pi}^{\text{e}}_{\text{H}}(\tau)}{\dd \tau}\dd{ \tau}+\int_0^t \mathcal{D}(t-\tau)\dfrac{\dd \bm{\Pi}^{\text{e}}_{\text{D}}(\tau)}{\dd \tau}\dd{ \tau}.
	\end{split}
\end{equation}
By following \cite{DePascalis2014}, we define:
\begin{equation}\label{eq:PiHD}
	\bm{\Pi}^{\text{e}}_{\text{H}}=J \textbf{F}^{-1}\left(\dfrac{1}{3}\text{tr}(\textbf{T}^{\text{e}})\textbf{I}\right)\textbf{F}^{-\text{T}}\qquad \text{and}\qquad \bm{\Pi}^{\text{e}}_{\text{D}}=J \textbf{F}^{-1}\left(\text{dev}(\textbf{T}^{\text{e}})\right)\textbf{F}^{-\text{T}},
\end{equation}
so that $\bm{\Pi}^{\text{e}}=\bm{\Pi}^{\text{e}}_{\text{H}}+\bm{\Pi}^{\text{e}}_{\text{D}}$.
Note that the relaxation functions $\mathcal{H}(t)$ and $\mathcal{D}(t)$ are associated to the Piola transformations of the hydrostatic and deviatoric parts of the Cauchy stress, $\bm{\Pi}^{\text{e}}_{\text{H}}$ and $\bm{\Pi}^{\text{e}}_{\text{D}}$, respectively. Moreover, by comparing \Eqs{eq:lin-sigma} and \eqref{QLV-initial} we see that they are both written with respect to the same tensorial relaxation function $\mathbb{G}(t)$. Therefore, the components $\mathcal{H}(t)$ and $\mathcal{D}(t)$ can be determined by performing step-strain tests in the linear regime. Upon a closer inspection of \Eq{eq:VEbases1} we can also note that $\mathcal{H}(t)=\kappa(t)/\kappa_0$ and $\mathcal{D}(t)=\mu(t)/\mu_0$ are the non-dimensional version of the relaxation functions $\kappa(t)$ and $\mu(t)$, where $\mu_0=\mu(0)$ and $\kappa_0=\kappa(0)$ are the instantaneous elastic bulk and shear modulus, respectively.\\
Finally, the Cauchy stress tensor follows from applying the transformation \mbox{$\textbf{T}=\!J^{-1} \textbf{F}\bm{\Pi}\textbf{F}^{\text{T}}$} to \Eq{eq:QLV-final} and is given by:
\begin{equation}\label{eq:QLV-cauchy}
	\textbf{T}(t)=J^{-1}(t)\textbf{F}(t)\left(\int_0^t \mathcal{H}(t-\tau)\dfrac{\dd \bm{\Pi}^{\text{e}}_{\text{H}}(\tau)}{\dd \tau}\dd{ \tau}+\int_0^t \mathcal{D}(t-\tau)\dfrac{\dd \bm{\Pi}^{\text{e}}_{\text{D}}(\tau)}{\dd \tau}\dd{ \tau}\right)\textbf{F}^{\text{T}}(t).
\end{equation}
\Eq{eq:QLV-final} is the QLV form of the constitutive equation for an isotropic compressible viscoelastic material. In the incompressible limit $J\!\rightarrow\!1$ and $\kappa(t)\!\rightarrow\!\kappa_0\!\rightarrow\!\infty$, $\forall t$, therefore \Eq{eq:QLV-cauchy} reduces to the following form:
\begin{equation}\label{eq:QLV-cauchy-inc}
	\textbf{T}(t)=\textbf{F}(t)\left(\int_0^t \mathcal{D}(t-\tau)\dfrac{\dd \bm{\Pi}^{\text{e}}_{\text{D}}(\tau)}{\dd \tau}\dd{ \tau}\right)\textbf{F}^{\text{T}}(t)-p(t)\textbf{I}
\end{equation}
where the Lagrange multiplier $p(t)$ is given by:
\begin{equation}\label{eq:p-QLV}
	p(t)=\lim_{k(t)\rightarrow\infty}\lim_{J\rightarrow 1}\left(J^{-1}(t)\textbf{F}(t)\left(\int_0^t \dfrac{\kappa(t-\tau)}{\kappa_0}\dfrac{\dd \bm{\Pi}^{\text{e}}_{\text{H}}(\tau)}{\dd \tau}\dd {\tau}\right)\textbf{F}^{\text{T}}(t)\right).
\end{equation}
\Eq{eq:QLV-cauchy-inc} is the QLV form of the constitutive equation for an isotropic incompressible material. \\
In the next section we consider the simple torsion of a solid cylinder. We derive the analytical expressions of the torque and the normal force required to twist the cylinder, both in the linear and the large deformations regime. We then derive the analytical expression for the relaxation curves of the torque and the normal force in two experimental scenarios: the step-strain test and the ramp test.

\section{Simple Torsion}\label{sec:torsion}
In this section we consider the problem of simple torsion of a solid cylinder.
We start by defining the coordinates of the cylinder in the reference configuration $\mathcal{B}_0$ and in the deformed configuration $\mathcal{B}(t)$ as $\{R,\Theta,Z\}$ and $\{r(t),\theta(t),z(t)\}$, respectively. We assume that the deformation starts at time $t\!=\!0$ and take the reference configuration as the initial configuration $\mathcal{B}_0\!=\!\mathcal{B}(0)$. The displacement vectors $\textbf{X}$ and $\textbf{x}(t)$ in $\mathcal{B}_0$ and $\mathcal{B}(t)$, respectively, are defined with respect to the bases $\{\textbf{E}_R,\textbf{E}_{\Theta},\textbf{E}_Z\}$ and $\{\textbf{e}_r,\textbf{e}_{\theta},\textbf{e}_z\}$, so that $\textbf{X}=R\textbf{E}_R+\Theta\textbf{E}_{\Theta}+Z\textbf{E}_Z$ and $\textbf{x}(t)=r(t)\textbf{e}_r+\theta(t)\textbf{e}_{\theta}+z(t)\textbf{e}_z$. The deformation can then be written as follows:
\begin{equation}\label{eq:def-tor}
	r(t)=R,\qquad \theta(t)=\Theta+\phi(t)Z,\qquad z(t)=Z,
\end{equation}
where $\phi(t)=\alpha(t)/l$ is the amount of twist experienced by the cylinder at time $t$, defined as the angle of rotation $\alpha(t)$ per unit length. $l$ is the length of the cylinder which remains constant at all times. The strain $\gamma(r,t)$, a non-dimensional measure of the deformation is:
\begin{equation}\label{gamma}
	\gamma(r,t)=\dfrac{r\alpha(t)}{l}=r\phi(t).
\end{equation}
The deformation gradient $\textbf{F}(r,t)=\dfrac{\partial\textbf{x}(t)}{\partial\textbf{X}}$ is given by:
\begin{equation}\label{eq:F-tor}
	\textbf{F}(r,t)=\left(\begin{array}{ccc}
			1 & 0 & 0           \\
			0 & 1 & r \,\phi(t) \\
			0 & 0 & 1           \\
		\end{array}\right)
\end{equation}
and the left Cauchy-Green tensor $\textbf{B}(r,t)=\textbf{F}(r,t)\textbf{F}(r,t)^{\text{T}}$ and its inverse are given by:
\begin{equation}\label{eq:B-tor}
	\textbf{B}(r,t)=\left(\begin{array}{ccc}
			1 & 0              & 0        \\
			0 & 1+r^2\phi^2(t) & r\phi(t) \\
			0 & r\phi(t)       & 1        \\
		\end{array}\right)\qquad \text{and}\qquad
	\textbf{B}(r,t)^{-1}=\left(\begin{array}{ccc}
			1 & 0         & 0              \\
			0 & 1         & -r\phi(t)      \\
			0 & -r\phi(t) & 1+r^2\phi^2(t) \\
		\end{array}\right).
\end{equation}
Note that the deformation gradient depends on the spatial variable $r$, \ie the deformation is non-homogeneous and the stress distribution will depend on the radial position as well.\\
The principal stretches and the principal directions associated to the torsion deformation are the eigenvalues and the eigenvectors of the tensor $\textbf{B}$, respectively. Upon diagonalising $\textbf{B}$, we find that the principal stretches are given by:
\begin{equation}\label{eq:lamPrinc}
	\lambda_1=1, \quad \lambda_{2,3}(r,t)=\sqrt{ 1+\frac{\gamma(r,t)}{2}\left(\gamma(r,t)\pm\sqrt{\gamma^2(r,t)+4} \right)}.
\end{equation}
$\lambda_2$ and $\lambda_3$ are the greatest and the smallest stretch respectively and the associated eigenvectors are the directions where $\lambda_2$ and $\lambda_3$ occur.
Note that both $\lambda_2$ and $\lambda_3$ depend on the spatial variable $r$. Moreover, $\lambda_2$ is maximum at the outer surface $r\!=\!r_o$. It is useful to define the strain $\gamma_o(t)$ as the strain at the outer surface of the cylinder at time $t$:
\begin{equation}\label{eq:gamma0}
	\gamma_o(t)=\gamma(r_o,t)=r_o\phi(t)
\end{equation}
so that:
\begin{equation}\label{eq:maxlambda2}
	\max_r \lambda_2(r,t)=\lambda_2(r_o,t)=\sqrt{ 1+\frac{\gamma_o(t)}{2}\left(\gamma_o(t)+\sqrt{\gamma_o^2(t)+4} \right)}
\end{equation}

\begin{figure}[t!]
	\centering
	\subfigure[a][]{\includegraphics[scale=0.3]{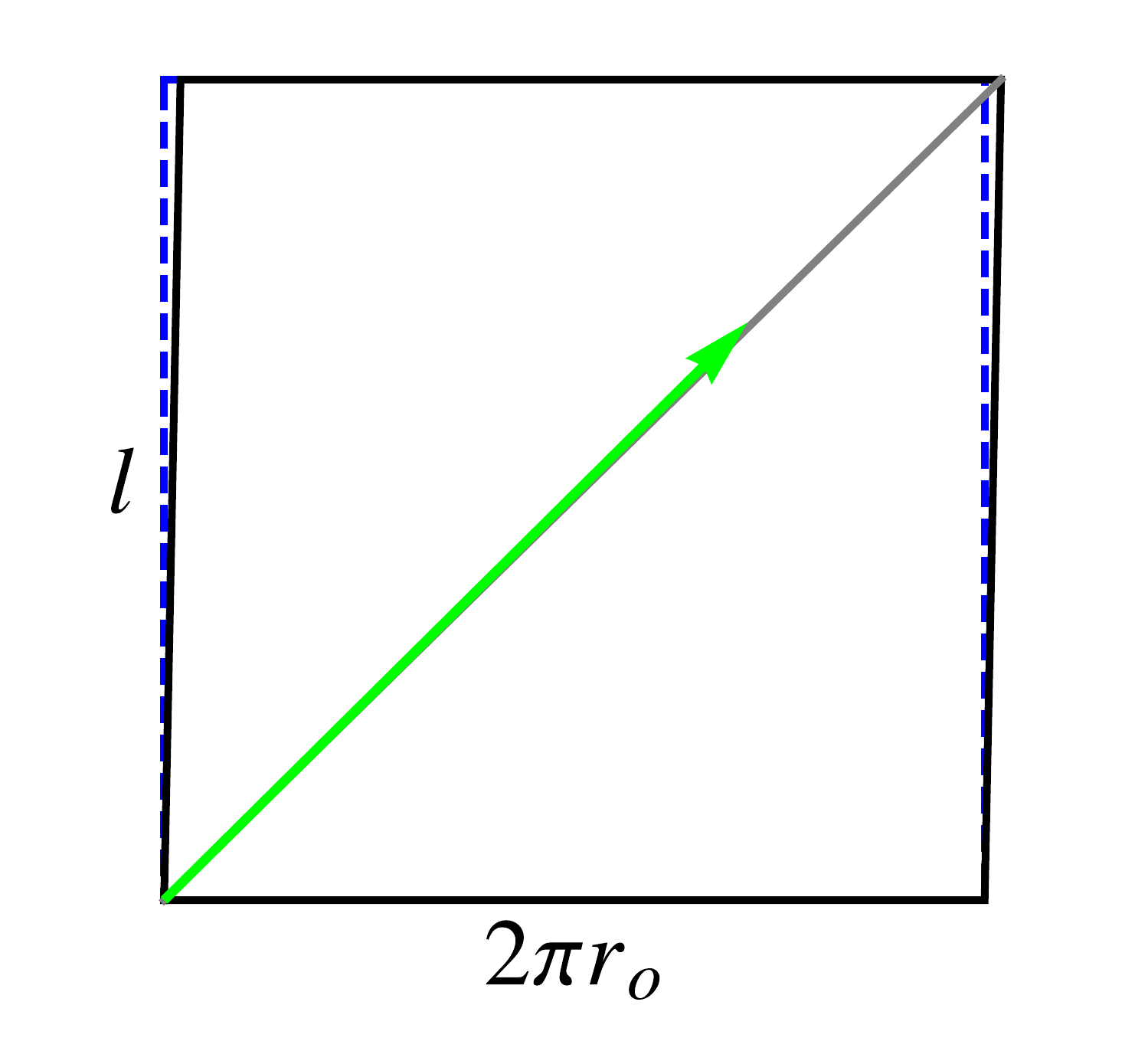}}
	\subfigure[b][]{\includegraphics[scale=0.3]{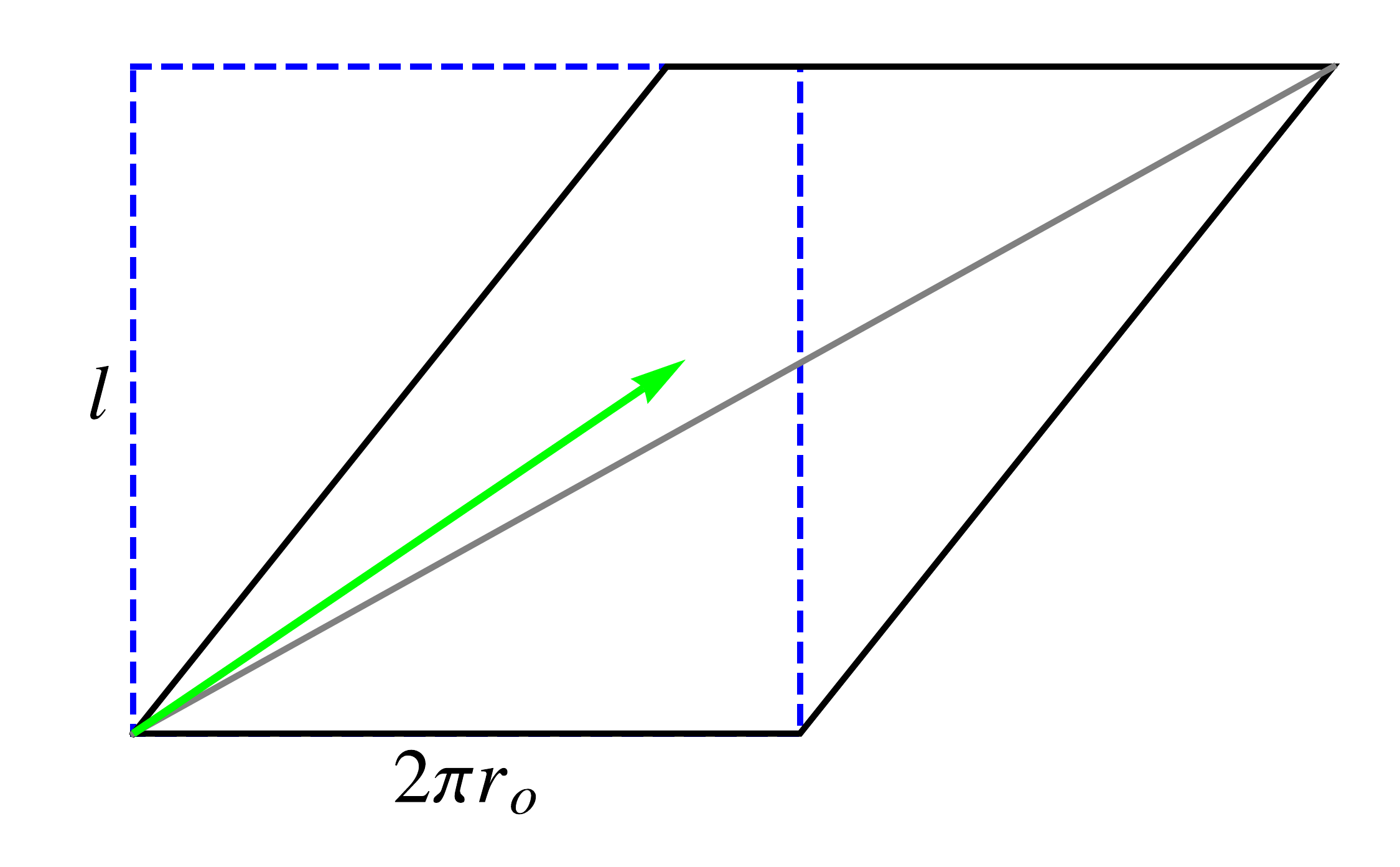}}
	\caption{Sketch of the lateral surface of a cylinder with length $l$ and radius $r_o$ twisted with $\gamma_o=0.02$ (a) and $\gamma_o=0.8$ (b). The green arrow shows the principal direction associated to the maximum stretch $\lambda_2$ in \Eq{eq:lamPrinc}. The dashed and the solid lines represent the undeformed and the deformed cylinder respectively.}\label{fig:principal}
\end{figure}

In \Fig{fig:principal} we show the principal direction (green arrow) associated to the maximum stretch (\ie $\lambda_2(r_o,t)$) on the external surface of the cylinder, when the cylinder experiences a strain $\gamma_o(t)=0.02$ (\Fig[a]{fig:principal}) and $\gamma_o(t)=0.8$ (\Fig[b]{fig:principal}). When $\gamma_o(t)\ll1$, \ie in the small deformation regime, the principal direction is aligned with the diagonal of the rectangle and the maximum stretch $\lambda_2(r_o,t)$ can be approximated by the following expansion:
\begin{equation}\label{eq:maxl2}
	\lambda_2(r_o,t)=1+\dfrac{\gamma_o(t)}{2}+\mathcal{O}(\gamma_o^2(t)),
\end{equation}
which recovers the relation between stretch $\lambda$ used for large deformations and the infinitesimal strain $\varepsilon$ used in small deformations $\lambda=\varepsilon+1$, with $\varepsilon=\gamma/2$.\\
However, \Fig[b]{fig:principal} shows that in the large deformation regime, the principal direction associated to the maximum stretch $\lambda_2(r_o,t)$ is not aligned with the diagonal and $\lambda_2(r_o,t)$ is given by \Eq{eq:maxlambda2}.

We can now write the governing equations for the simple torsion of a solid cylinder. Assuming that the inertia is negligible, the equilibrium equations at any time $t>0$ are given by:
\begin{equation}\label{eq:goveqs}
	\left\{\begin{split}
		\text{div}\textbf{T}(r,t)&=\boldsymbol{0}\\
		T_{rr}(r,t)&=0\qquad\text{at}\qquad r=r_o
	\end{split}\right.
\end{equation}
where $r_o$ is the outer radius of the cylinder and the last equation follows from imposing that the lateral surface of the cylinder is free of traction at any time $t$. The operator $\text{div}$ is the divergence operator (in cylindrical coordinates), see \cite{OgdenBook} for details.
In writing \Eq{eq:goveqs} we have implicitly neglected inertial forces. For a discussion on the validity of this assumption we refer to \cite{Balbi2019visco,SaccoInertia}.

The Cauchy stress $\textbf{T}(r,t)$ is given by the constitutive equation \eqref{eq:QLV-cauchy-inc}. Given the form of the tensors $\F(r,t)$ and $\textbf{B}(r,t)$ from \Eqs{eq:F-tor} and \eqref{eq:B-tor}, the Cauchy stress tensor will have components $T_{r\theta}(r,t)\!=\!T_{\theta r}(r,t)\!=\!T_{rz}(r,t)\!=\!T_{zr}(r,t)\!=\!0$, $\forall t$. The remaining non-zero components are:
\begin{equation}\label{sig-p}
	\begin{split}
		T_{rr}(r,t)&=\int_0^t\mathcal{D}(t-\tau)\dfrac{\partial}{\partial \tau}\PieDxx(r,\tau)\dd{\tau}-p(r,\theta,z,t)\\
		T_{\theta\theta}(r,t)&=\int_0^t\mathcal{D}(t-\tau)\dfrac{\partial}{\partial\tau}\PieDyy(r,\tau)\dd{\tau}+2r\phi(t)\int_0^t\mathcal{D}(t-\tau)\dfrac{\partial}{\partial\tau}\PieDyz(r,\tau)\dd{\tau}
		\\&+r^2\phi(t)^2\int_0^t\mathcal{D}(t-\tau)\dfrac{\partial}{\partial\tau}\PieDzz(r,\tau)\dd{\tau}-p(r,\theta,z,t)\\
		T_{zz}(r,t)&=\int_0^t\mathcal{D}(t-\tau)\dfrac{\partial}{\partial\tau}\PieDzz(r,\tau)\dd{\tau}-p(r,\theta,z,t)\\
		T_{\theta z}(r,t)&=\int_0^t\mathcal{D}(t-\tau)\dfrac{\partial}{\partial\tau}\PieDyz(r,\tau)\dd{\tau}+r\phi(t)\int_0^t\mathcal{D}(t-\tau)\dfrac{\dd}{\partial\tau}\PieDzz(r,\tau)\dd{\tau}\\
	\end{split}
\end{equation}
Therefore, the governing equations reduce to:
\begin{equation}\label{div-reduced}
	\left\{
	\begin{split}
		&\dfrac{\partial T_{rr}(r,t)}{\partial r}+\dfrac{T_{rr}(r,t)-T_{\theta\theta}(r,t)}{r}=0\\
		&\dfrac{\partial T_{\theta\theta}(r,t)}{\partial \theta}=0\\
		&\dfrac{\partial T_{zz}(r,t)}{\partial z}=0
	\end{split}
	\right.
\end{equation}
From the last two equations in \Eqs{div-reduced} we can conclude that the Lagrange multiplier $p$ only depends on the spatial variable $r$ and at any time $t$ the governing problem reduces to a single Ordinary Differential Equation (ODE) in the argument $r$:
\begin{equation}\label{div-eq}
	\left\{\begin{split}
		\dfrac{\dd T_{rr}}{\dd r}+\dfrac{T_{rr}-T_{\theta\theta}}{r}&=0\\
		T_{rr}&=0\qquad\text{at}\qquad r=r_o
	\end{split}\right.
\end{equation}
Now, we restrict our attention to soft tissues whose elastic behaviour can be considered hyperelastic. For such tissues a strain energy function $W$ can be defined. Here, we choose $W$ in the form of the Mooney-Rivlin model:
\begin{equation}\label{eq:MR}
	W(I_1,I_2)=(\frac{\mu_0}{2}-c_2)(I_1-3)+c_2(I_2-3),
\end{equation}
where $I_1\!=\!\tr \B$ and $I_2=\nicefrac{1}{2}\left( (\tr\B)^2-\tr\B^2\right) $ are the first and second invariants of the tensor $\B$, respectively, $\mu_0$ is the infinitesimal shear modulus and $c_2$ is the second Mooney-Rivlin parameter. The Neo-Hookean model is recovered by setting $c_2=0$.
The choice for $W$ is motivated by many experimental observations on soft tissues. In particular, it has been observed that the brain behaves as a Mooney-Rivlin material in torsion \cite{Balbi2019} and in simple shear \cite{RachidShear2013}. Moreover, the Mooney-Rivlin model has the key feature of predicting a linear elastic response in torsion, \ie the torque required to twist the cylinder depends linearly on the strain.
The elastic Cauchy stress $\textbf{T}^{\text{e}}$ for an incompressible material is given by the following relation:
\begin{equation}\label{eq:cauchy-el}
	\textbf{T}^{\text{e}}=2 W_1\textbf{B}-2 W_2 \textbf{B}^{-1}-p^{\text{e}} \textbf{I},
\end{equation}
where $W_i=\partial W/\partial I_i$, $i=\{1,2\}$ and $p^{\text{e}}$ is the elastic Lagrange multiplier \cite{OgdenBook}.
By combining \Eqs{eq:MR}, \eqref{eq:cauchy-el} and \eqref{eq:PiHD} we can calculate the components of the tensor $\PieD$:
\begin{equation}\label{PieD}
	\begin{split}
		\PieDxx(r,t)&=\nicefrac{1}{3}(4 c_2-\mu_0)r^2\phi^2(t),\\
		\PieDyy(r,t)&=-\nicefrac{2}{3}(c_2+2\mu_0)r^2\phi^2(t)-\nicefrac{1}{3}(2c_2+\mu_0)r^4\phi^4(t),\\
		\PieDzz(r,t)&=-\nicefrac{1}{3}(2 c_2+\mu_0)r^2\phi^2(t),\\
		\PieDyz(r,t)&=\mu_0 r \phi(t)+\nicefrac{1}{3}(2 c_2+\mu_0)r^3\phi^3(t).
	\end{split}
\end{equation}
By combining \Eqs{sig-p} and \Eqs{PieD} and substituting into \Eq{div-eq} we obtain an ODE for the variable $p$:
\begin{equation}\label{solve-p}
	\begin{split}
		\dfrac{\dd p}{\dd r}&=\dfrac{r}{3}  (14 c_2 + \mu_0) \int^t_0\mathcal{D}(t-\tau)\dfrac{\dd}{\dd\tau}\phi^2(\tau)\dd \tau + \dfrac{r^3}{3} (2 c_2 + \mu_0) \int^t_0\mathcal{D}(t-\tau)\dfrac{\dd}{\dd\tau}\phi^4(\tau) \dd{\tau}\\
		&-\phi(t) \left(2 r \mu \int^t_0\mathcal{D}(t-\tau)\dfrac{\dd}{\dd\tau}\phi(\tau)\dd{\tau} +
		\dfrac{2}{3} r^3 (2 c_2 + \mu) \int^t_0\mathcal{D}(t-\tau)\dfrac{\dd}{\dd\tau}\phi^3(\tau)\dd{\tau}\right) \\
		&+  \phi^2(t) \dfrac{r^3}{3} (2 c_2 + \mu_0) \int^t_0\mathcal{D}(t-\tau)\dfrac{\dd}{\dd\tau}\phi^2(\tau)\dd{\tau}
	\end{split}
\end{equation}
with the initial condition:
\begin{equation}\label{cond-p}
	p= \dfrac{r_o^2}{3}(4 c_2 - \mu_0)\int_0^t\mathcal{D}(t-\tau)\dfrac{\dd}{\dd\tau}\phi^2(\tau)\dd{\tau}\qquad\text{at}\,\,r=r_o
\end{equation}
whose solution is:
\begin{equation}\label{p}
	\begin{split}
		p&= \dfrac{1}{6}\left((14 c_2 +\mu_0)r^2 -  3(2 c_2  +   \mu_0)r_o^2\right)  \int^t_0\mathcal{D}(t-\tau)\dfrac{\dd}{\dd\tau}\phi^2(\tau)\dd{\tau} \\
		&+\dfrac{1}{12} (2 c_2 + \mu_0) (r^4 - r_o^4) \int^t_0\mathcal{D}(t-\tau)\dfrac{\dd}{\dd\tau}\phi^4(\tau)\dd{\tau}\\
		&- \left( \mu_0 (r^2 - r_o^2) \int^t_0\mathcal{D}(t-\tau)\dfrac{\dd}{\dd\tau}\phi(\tau)\dd{\tau}\right.\\
		&\left.+
		\dfrac{1}{6} (2 c_2 + \mu_0) (r^4 - r_o^4)  \int^t_0\mathcal{D}(t-\tau)\dfrac{\dd}{\dd\tau}\phi^3(\tau)\dd{\tau}\right) \phi(t)\\
		&+\left(\dfrac{1}{12} (2 c_2 + \mu_0) (r^4 - r_o^4)   \int^t_0\mathcal{D}(t-\tau)\dfrac{\dd}{\dd\tau}\phi^2(\tau)\dd{\tau}\right)\phi^2(t)
	\end{split}
\end{equation}
Finally, the components of the stress $\textbf{T}(r,t)$ can be obtained by substituting \Eq{p} into \Eqs{sig-p} to fully determine the final stress distribution in the cylinder.

The torque $T(t)$ required to twist the cylinder can be computed as:
\begin{equation}\label{key}
	T(t)=\int_0^{2\pi}\int_0^{r_o}T_{\theta z}(r,t)r^2\dd{r}\dd{\theta},
\end{equation}
and the normal force $N$ necessary to keep to cylinder length constant reads:
\begin{equation}\label{key}
	N(t)=\int_0^{2\pi}\int_0^{r_o}T_{zz}(r,t)r\dd{r}\dd{\theta}.
\end{equation}

The components $T_{\theta z}(r,t)$ and $T_{zz}(r,t)$ are given by \Eqs{sig-p} upon substituting \Eqs{PieD} and \eqref{p}. The final expressions for the torque and the normal force read:
\begin{equation}\label{qlv-torque}
	\begin{split}
		T(t)&=\dfrac{\pi}{2}  \mu_0 r_o^4\int^t_0\mathcal{D}(t-\tau)\dfrac{\dd}{\dd\tau}\phi(\tau)\dd{\tau}\\
		&+\dfrac{\pi}{9} (2c_2+\mu_0)r_o^6\left(\int^t_0\mathcal{D}(t-\tau)\dfrac{\dd}{\dd\tau}\phi^3(\tau)\dd{\tau}-\phi(t)\int^t_0\mathcal{D}(t-\tau)\dfrac{\dd}{\dd\tau}\phi^2(\tau)\dd{\tau}\right)
	\end{split}
\end{equation}
and
\begin{equation}\label{qlv-force}
	\begin{split}
		N(t)&=-\dfrac{\pi}{2} \mu_0  r_o^4 \phi(t)\int^t_0\mathcal{D}(t-\tau)\dfrac{\dd}{\dd\tau}\phi(\tau)\dd{\tau}-\dfrac{\pi}{4} (2 c_2-\mu_0) r_o^4\int^t_0\mathcal{D}(t-\tau)\dfrac{\dd}{\dd\tau}\phi^2(\tau)\dd{\tau}\\
		&+\dfrac{\pi}{18} (2 c_2 + \mu_0)r_o^6 \left(\int^t_0\mathcal{D}(t-\tau)\dfrac{\dd}{\dd\tau}\phi^2(\tau)\dd{\tau}\right)\phi^2(t)\\
		&-\dfrac{\pi}{9} (2 c_2 + \mu_0) r_o^6 \left(\int^t_0\mathcal{D}(t-\tau)\dfrac{\dd}{\dd\tau}\phi^3(\tau)\dd{\tau}\right)\phi(t)\\
		&+\dfrac{\pi}{18} (2 c_2 + \mu_0) r_o^6 \int^t_0\mathcal{D}(t-\tau)\dfrac{\dd}{\dd\tau}\phi^4(\tau)\dd{\tau},
	\end{split}
\end{equation}
respectively.\\
Note that \Eqs{qlv-torque} and \eqref{qlv-force} are written with respect to the twist $\phi(t)$ which is a dimensional measure of the deformation. In view of comparing the predictions of the QLV theory with those of the linear theory, it is useful to rewrite \eqref{qlv-torque} and \eqref{qlv-force} in terms of the strain $\gamma_o(t)$, defined in \Eq{eq:gamma0}, which is the strain at the outer surface of the cylinder and is a non-dimensional measure of the deformation.
Then, \Eqs{qlv-torque} and \eqref{qlv-force} rewrite as follows:
\begin{equation}\label{eq:qlv-torque-gamma}
	\begin{split}
		T(t)&=\dfrac{\pi}{2} r_o^3\int^t_0\mu(t-\tau)\dfrac{\dd}{\dd\tau}\gamma_o(\tau)\dd{\tau}\\
		&+\dfrac{\pi}{9} (\dfrac{2c_2}{\mu_0}+1)r_o^3\left(\int^t_0\mu(t-\tau)\dfrac{\dd}{\dd\tau}\gamma_o^3(\tau)\dd{\tau}-\gamma_o(t)\int^t_0\mu(t-\tau)\dfrac{\dd}{\dd\tau}\gamma_o^2(\tau)\dd{\tau}\right)
	\end{split}
\end{equation}
and
\begin{equation}\label{eq:qlv-force-gamma}
	\begin{split}
		N(t)&=-\dfrac{\pi}{2}  r_o^2 \gamma_o(t)\int^t_0\mu(t-\tau)\dfrac{\dd}{\dd\tau}\gamma_o(\tau)\dd{\tau}-\dfrac{\pi}{4} (\dfrac{2 c_2}{\mu_0}-1) r_o^2\int^t_0\mu(t-\tau)\dfrac{\dd}{\dd\tau}\gamma_o^2(\tau)\dd{\tau}\\
		&+\dfrac{\pi}{18} (\dfrac{2 c_2}{\mu_0} + 1)r_o^2 \left(\int^t_0\mu(t-\tau)\dfrac{\dd}{\dd\tau}\gamma_o^2(\tau)\dd{\tau}\right)\gamma_o^2(t)\\
		&-\dfrac{\pi}{9} (\dfrac{2 c_2}{\mu_0} + 1) r_o^2 \left(\int^t_0\mu(t-\tau)\dfrac{\dd}{\dd\tau}\gamma_o^3(\tau)\dd{\tau}\right)\gamma_o(t)\\
		&+\dfrac{\pi}{18} (\dfrac{2 c_2}{\mu_0} + 1) r_o^2 \int^t_0\mu(t-\tau)\dfrac{\dd}{\dd\tau}\gamma_o^4(\tau)\dd{\tau}
	\end{split}
\end{equation}
where we have used the connection $\mathcal{D}(t)\!=\!\mu(t)/\mu_0$.\\
In the next section we will use \Eqs{eq:qlv-torque-gamma} and \eqref{eq:qlv-force-gamma} to calculate the predictions of the QLV model in the experimental scenarios of a step-strain test and a ramp test.

\section{Results}\label{sec:results}
In view of predicting the relaxation behaviour of a tissue in simple torsion, we consider two scenarios which are important from the experimental viewpoint: the step-strain test and the ramp test. By using \Eqs{eq:qlv-torque-gamma} and \eqref{eq:qlv-force-gamma}, we then derive the analytical expressions of the relaxation curves for the torque and the normal force.

\subsection{Small Deformations}\label{sec:lin-tors}
We start by considering the torsion of a cylindrical tissue in the small deformation regime and we calculate the torque and the normal force predicted by the linear viscoelastic theory presented in \Sec{sec:Linear}. We assume that the viscoelastic response of the tissue can be modelled as that of a generalised Maxwell model, \ie a system of spring and dash-pots arranged as shown in \Fig{fig:gen-max}.
For small deformations, the only non-zero component of the infinitesimal strain tensor $\bm{\varepsilon}$ is $\varepsilon_{\theta z}=\gamma(r,t)/2$, thus it follows that $\text{dev}\bm{\varepsilon}=\bm{\varepsilon}$ and the constitutive equation \eqref{eq:linconst} reduces to the following equation:
\begin{equation}\label{eq:sigmatorsionlin}
	\sigma_{\theta z}(r,t)=\int_{-\infty}^t\mu(t-\tau)\dfrac{\dd}{\dd \tau}\gamma(r,\tau)\dd{\tau}.
\end{equation}
The governing equations \eqref{eq:goveqs} are automatically satisfied.
\Eq{eq:sigmatorsionlin} is a one-dimensional equation in the same form of \Eq{eq:sigma1dof}, where $\mu(t)$ is now the relaxation function.
According to the formula $T_{\text{lin}}=2\pi\int_0^{r_o}\sigma_{\theta z}r^2\dd r$, the torque is then given by:
\begin{equation}\label{eq:torque-lin}
	T_{\text{lin}}(t)=\dfrac{\pi}{2}r_o^3\int_{-\infty}^t\mu(t-\tau)\dfrac{\dd}{\dd \tau}\gamma_o(\tau)\dd{\tau},
\end{equation}
where $\gamma_o(t)$ is the shear evaluated at the outer radius $r_o$, see \Eq{eq:gamma0}.
Note that \Eq{eq:torque-lin} is the linearised version of \Eq{eq:qlv-torque-gamma} for $\gamma_o\ll1$.

Following \Sec{sec:rel-fun}, we can calculate the analytical expressions of the relaxation curve for the torque in response to a step and a ramp input, respectively. \\
For the step-strain test, we consider the following form for the strain $\gamma_o(t)$:
\begin{equation}\label{eq:gamma2}
	\gamma_o(t)=\frac{r_o}{l}\alpha_0 H(t)=\gamma_{o,0}H(t),
\end{equation}
where $\gamma_{o,0}=\frac{r_o}{l}\alpha_0$ is the amplitude of the step and $H(t)$ is the Heaviside function as defined in \Sec{sec:rel-fun}.
By substituting \Eq{eq:gamma2} into \Eq{eq:torque-lin} and upon integrating, we obtain:
\begin{equation}\label{eq:Tlin}
	T^{\text{lin}}(t)=\frac{1}{2}\pi r_o^3 \mu_{\text{step}}(t) \gamma_{o,0}
\end{equation}
where the relaxation function $\mu_{\text{step}}(t)$ is the following Prony series:
\begin{equation}\label{eq:pronymu}
	\mu_{\text{step}}(t)=\dfrac{2T^{\text{lin}}(t)}{\pi r_o^3\gamma_{o,0}}=\mu_{\infty}+\sum_{i=1}^{n}\mu_i e^{-\frac{t}{\tau_i}}.
\end{equation}
The parameters $\mu_i$ in \Eq{eq:pronymu} are the shear moduli of the $n$ branches of the generalized Maxwell model, see \Fig{fig:gen-max}, and $\tau_i=\frac{\eta_i}{\mu_i}$ are the associated relaxation times. The instantaneous shear modulus is obtained by evaluating the maximum of the relaxation function $\mu_{\text{step}}(t)$ in $t=0$, as follows:
\begin{equation}\label{eq:mu0}
	\mu_0=\mu_{\text{ramp}}(0)=\mu_{\infty}+\sum_{i=1}^{n}\mu_i .
\end{equation}
We now consider a strain history in the form of a ramp, as follows:
\begin{equation}\label{eq:ramp2}
	\gamma_o(t)=\frac{\gamma_{o,0}}{t^*}t-\frac{\gamma_{o,0}}{t^*}(t-t^*)H(t-t^*).
\end{equation}
where $t^*$ is the rising time of the ramp.
Substituting \Eq{eq:ramp2} into \Eq{eq:torque-lin} and computing the integral provides an analogous expression for the torque as \Eq{eq:Tlin}. The relaxation curve of the torque is now given by:
\begin{equation}\label{eq:rframp}
	\mu_{\text{ramp}}(t)=\frac{2 T^{\text{lin}}(t)}{\pi r_o^3 \gamma_{o,0}}=
	\mu_{\infty}+\frac{1}{t^*}\sum_{i=1}^{n}\tau_i \mu_i e^{-\frac{t}{\tau_i}}\left( e^{\frac{t^*}{\tau_i}} -1 \right), \qquad \text{for} \quad t>t^*.
\end{equation}
\Eq{eq:rframp} describes the relaxation curve of the torque in response to the ramp function in \Eq{eq:ramp2}. The right-hand side term in \Eq{eq:rframp} is the response of the generalized Maxwell system to a ramp input.
\Eqs{eq:pronymu} and \eqref{eq:rframp} can then be used to estimate the viscoelastic parameters $\mu_i,\mu_{\infty}$ and $\tau_i$ by fitting the data from a step-test and a ramp-test, respectively. The left-hand side of \Eq{eq:rframp} can be computed from the experimental data, \ie the measured torque $T_{\text{lin}}$ and the imposed strain $\gamma_{o,0}$ and the radius of the sample $r_o$. The right-hand side is the analytical expression to be fitted in order to estimate the viscoelastic parameters.\\
Note that the expression on the right-hand side of \Eq{eq:rframp} does not depend on the amount of shear $\gamma_{o,0}$, therefore \Eq{eq:rframp} will predict accurate results for those tissues that display the same relaxation response when subjected to different levels of strain. Moreover, we define the following non-dimensional function:
\begin{equation}\label{eq:maxTorLin}
	\tilde{\mu}_{0\text{ramp}}=\dfrac{\mu_{\text{ramp}}(t^*)-\mu_{\infty}}{\mu_0}=\frac{1}{t^*}\sum_{i=1}^{n}\dfrac{\tau_i \mu_i}{\mu_0} \left(1- e^{-\frac{t^*}{\tau_i}}  \right)=\sum_{i=1}^{n}\dfrac{\nu_i^{-1} \mu_i}{\mu_0} \left(1- e^{-\nu_i}  \right),
\end{equation}
which provides an estimate of the maximum value of the relaxation curve in \Eq{eq:rframp}.

In the next section, we derive the corresponding expressions for the relaxation curves of the torque and the normal force for the QLV model.

\subsection{Large Deformations}\label{sec:largedef}
In this section we derive the relaxation curves for the torque and the normal force for a step test and a ramp test in the large deformation regime.
We first address the ramp test scenario and then by taking the limit case $t^*\rightarrow 0$ we consider the step test scenario.

\subsubsection{Torque}
To compute the analytical expression of the relaxation curve for the torque, we substitute the strain history \eqref{eq:ramp2} and the relaxation function \eqref{eq:pronymu} into \Eq{eq:qlv-torque-gamma}. Upon integrating we obtain:
\begin{equation}\label{key}
	T(t,\gamma_{o,0})=\frac{1}{2}\pi r_o^3 \, \mu_{\text{ramp}}^{\text{QLV}}(t,\gamma_{o,0}) \,\gamma_{o,0}, \qquad \text{for} \quad t>t^*
\end{equation}
where:
\begin{equation}\label{eq:rf-qlv}
	\begin{split}
		\mu_{\text{ramp}}^{\text{QLV}}(t,\gamma_{o,0})=\mu_{\text{ramp}}(t)+\frac{2(1+\nicefrac{2c_2}{\mu_0})}{9t^{*3}}\sum_{i=1}^{n}\tau_i \mu_i e^{-\frac{t}{\tau_i}} &\Big(
		-2\tau_i(t^*+3\tau_i)\\
		&+e^{\frac{t^*}{\tau_i}}(t^{*2}-4t^*\tau_i+6\tau_i^2)
		\Big) \gamma_{o,0}^2,
	\end{split}
\end{equation}
which is valid for $t>t^*$. \Eq{eq:rf-qlv} shows that the relaxation curve for the torque predicted by the QLV model is the sum of the relaxation function $\mu_{\text{ramp}}(t)$ in \Eq{eq:rframp} and a term proportional to the square of the imposed strain $\gamma_{o,0}$. Therefore, in the limit $\gamma_{o,0}\ll 1$, \ie small deformation regime, \Eq{eq:rf-qlv} recovers the result in \eqref{eq:rframp}, which is the relaxation function predicted by the linear theory for a ramp test.\\
A less straightforward comment is that the quadratic term in \Eq{eq:rf-qlv} arises as a consequence of the time-dependent nature of the strain history. In other words, if the tissue is deformed infinitely fast (as it is the case for a perfect step-strain), \Eq{eq:rf-qlv} recovers the function $\mu_{\text{step}}(t)$ in \eqref{eq:pronymu}, without making the assumption of small deformations! By taking the limit for $t^*\rightarrow 0$ in \Eq{eq:rf-qlv} we indeed obtain:
\begin{equation}\label{eq:muQLV0}
	\lim_{t^*\rightarrow 0}\mu_{\text{ramp}}^{\text{QLV}}(t,\gamma_{o,0})= \mu_{\infty}+\sum_{i=1}^{n}\mu_i e^{-\frac{t}{\tau_i}}=\mu_{\text{step}}(t),
\end{equation}
which is the relaxation curve predicted by the linear theory for a step test.
Moreover, opposite to the linear viscoelastic theory, the relaxation curve of the torque predicted by the QLV model, \ie $\mu_{\text{ramp}}^{\text{QLV}}(t,\gamma_{o,0})$, depends on both the time and the level of strain $\gamma_{o,0}$. This is in agreement with the original assumption made by Fung when he first formulated the QLV theory \cite{Fung2013}.\\
With the aim of plotting the relaxation curve $\mu_{\text{ramp}}^{\text{QLV}}$ in \Eq{eq:rf-qlv}, we now consider a simplified version of the generalised Maxwell model with only one branch. This layout is also called Standard Linear Solid model, and it is the simplest arrangement of elements which is able to describe the behaviour of a viscoelastic solid.
Note that the QLV theory, despite taking into account large deformations, obeys the superposition principle in time. Therefore, adding more branches to the generalized Maxwell model will increase the accuracy of the model without producing mixed higher order terms as, for example, in the multiple integral formulation \cite{Lockett1972}. We can then set $n=1$ in \Eq{eq:rf-qlv} without loss of generality.
Furthermore, we set $c_2=c_1/2$ according to the observed values of $c_1$ and $c_2$ for brain tissues \cite{Balbi2019}.

\begin{figure}[htb]
	\centering
	\includegraphics[width=0.7\linewidth]{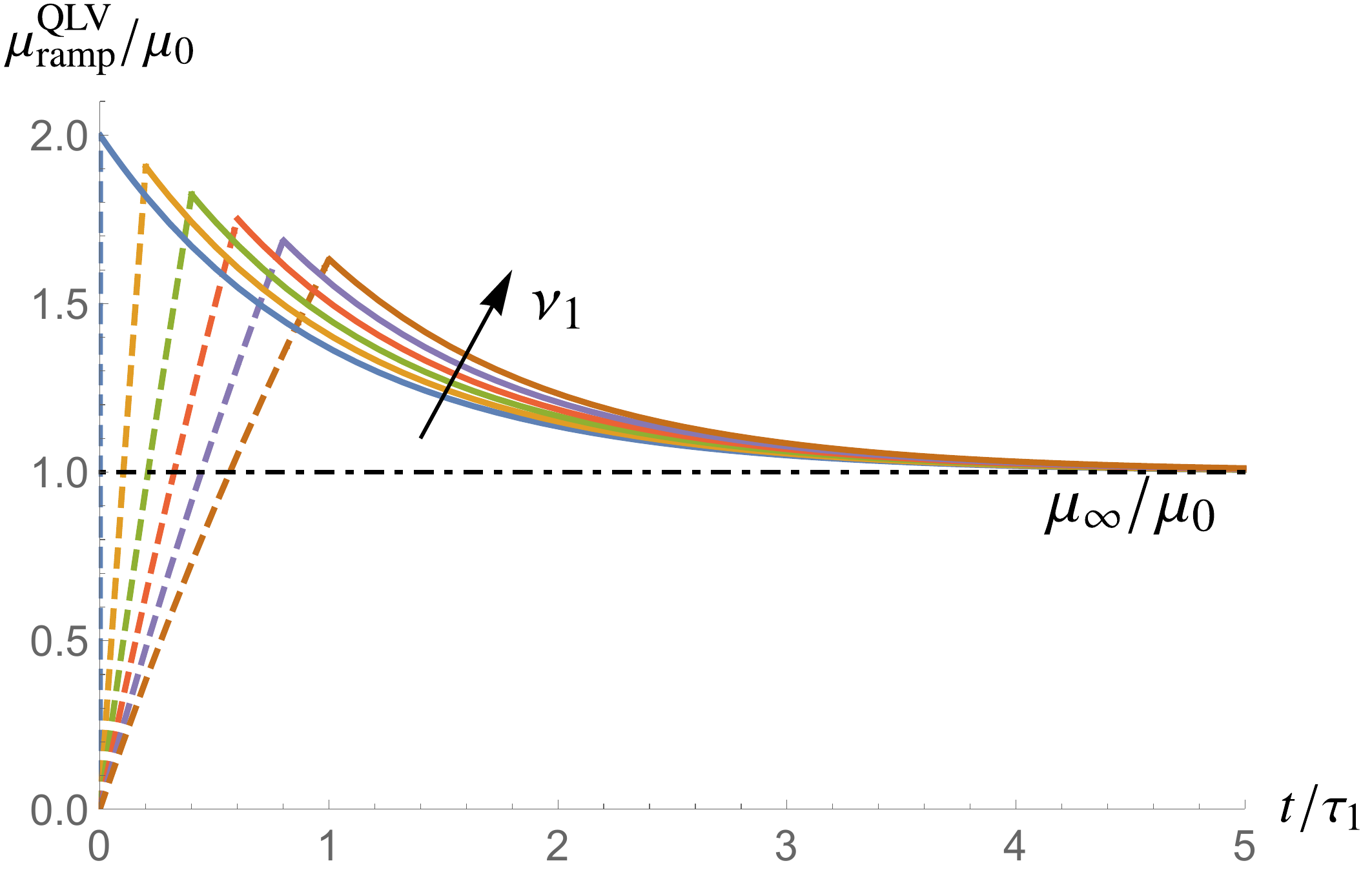}
	\caption{Effect of the rising time $t^*$ on the relaxation curve $\mu_{\text{ramp}}^{\text{QLV}}(t,\gamma_{o,0})/\mu_0$ in \Eq{eq:rf-qlv}. The parameter $\nu_1=t^*/\tau_1$ spans over $\{0,0.2,0.4,0.6,0.8,1\}$. The following parameters have been fixed: $n=1$, $\tau_1=1$, $\mu_1=\mu_{\infty}=1$, $c_2/\mu_0=2/3$ and the strain level $\gamma_{o,0}=0.02$.}	\label{fig:muqlvtnu}
\end{figure}

In \Fig{fig:muqlvtnu} we plot the relaxation curve $\mu_{\text{ramp}}^{\text{QLV}}(t,\gamma_{o,0})$ in \Eq{eq:rf-qlv} for six different ramp histories. To quantify the effect of the rising time on the profile of the relaxation curve we vary the parameter $\nu_1=t^*/\tau_1$. We note that \Eq{eq:rf-qlv} is valid for $t>t^*$. Experimentally, the ramp phase ($t<t^*$) is the noisy part of the data and cannot be used to perform the model fitting. Therefore, we plot the ramp phase of the curves in \Fig{fig:muqlvtnu} with dashed lines. The dashed lines can be obtained by integrating \Eq{eq:qlv-torque-gamma} with the strain history \eqref{eq:ramp2} in the time interval $0\le t<t^*$.\\
From \Fig{fig:muqlvtnu} we can conclude that the faster the ramp phase, the higher the peak of the relaxation curve. However, the limiting value of the curves as $t$ approaches $\infty$ is $\mu_{\infty}/\mu_0$ and is not affected by the rising time $t^*$.

\begin{figure}[htb]
	\centering
	\includegraphics[width=0.7\linewidth]{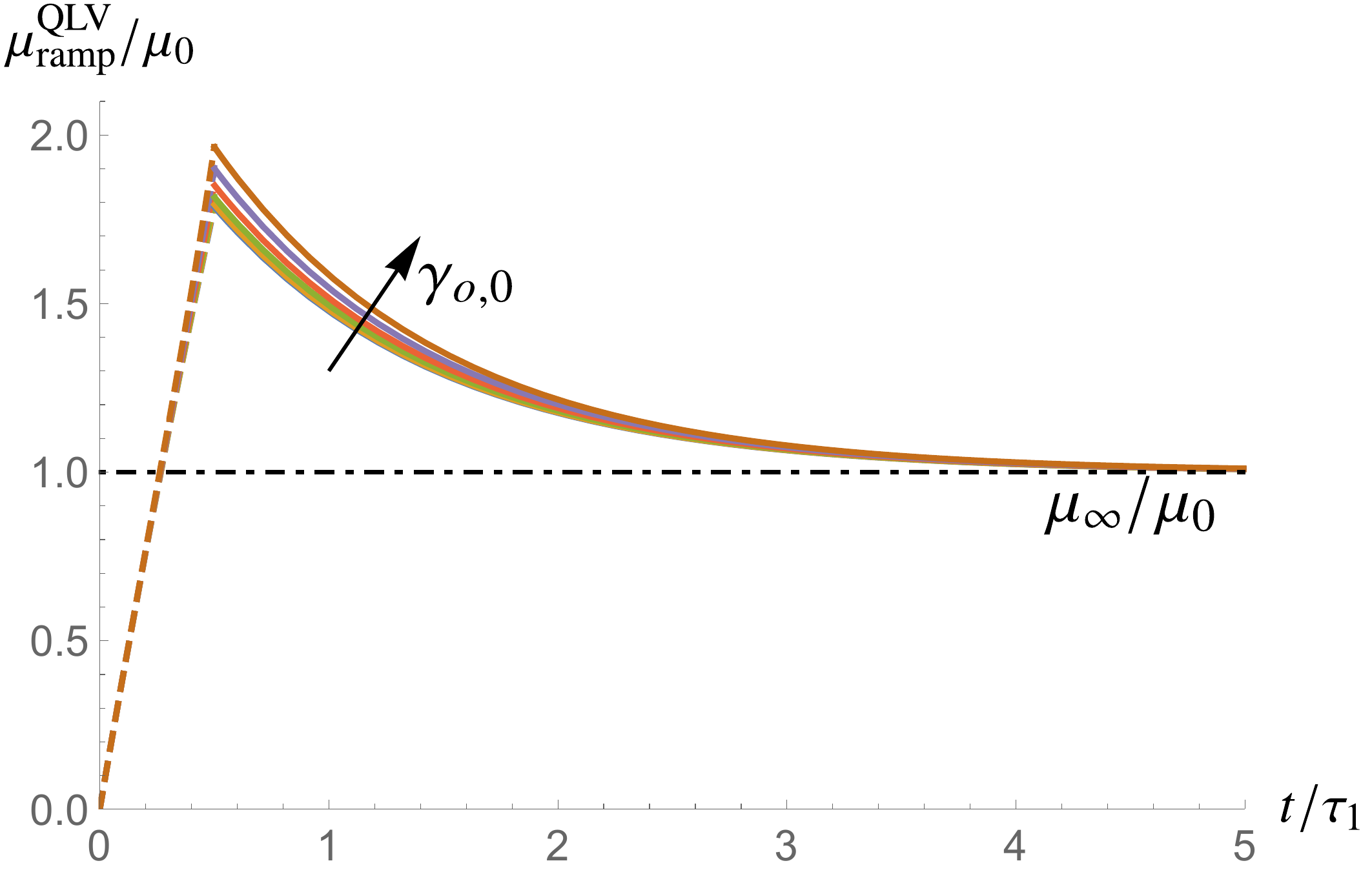}
	\caption{Effect of the final strain level $\gamma_{o,0}$ on the relaxation curve $\mu_{\text{ramp}}^{\text{QLV}}(t,\gamma_{o,0})/\mu_0$ in \Eq{eq:rf-qlv}. The strain $\gamma_{o,0}$ spans over $\{0.02,\pi/5,\pi/4,\pi/3,\pi/2,\pi\}$. The following parameters have been fixed: $n=1$, $\tau_1=1$, $\mu_1=\mu_{\infty}=1$, $c_2/\mu_0=2/3$ and the parameter $\nu_1=t^*/\tau_1=0.5$.}
	\label{fig:muqlvtgamma}
\end{figure}

To quantify the effect of the strain level reached at the end of the ramp phase, in \Fig{fig:muqlvtgamma} we plot \Eq{eq:rf-qlv} for different values of $\gamma_{o,0}$ at fixed $\nu_1=0.5$.

We now look at two limiting values of \Eq{eq:rf-qlv}, namely the long-term equilibrium and the instantaneous values of  $\mu_{\text{ramp}}^{\text{QLV}}(t,\gamma_{o,0})$.
First, by taking the limit for $t\rightarrow\infty$ we obtain:
\begin{equation}\label{eq:qlv-inf}
	\lim_{t\rightarrow \infty}	\mu_{\text{ramp}}^{\text{QLV}}(t,\gamma_{o,0})= \mu_{\infty},
\end{equation}
From \Eq{eq:qlv-inf} we observe that the equilibrium value of the relaxation curve of the torque predicted by the QLV model is not affected by the quadratic terms in $\gamma_{o,0}$. In other words, after an infinite time, \ie at the equilibrium, the response of the tissue is dominated by the long term shear modulus $\mu_{\infty}$. Therefore, time and its non-linear effect through the quadratic terms in \Eq{eq:rf-qlv} have no influence on the long-term modulus. This effect is also observable from the horizontal asymptotes of Figures~\ref{fig:muqlvtnu} and \ref{fig:muqlvtgamma}.

Then, we look at the instantaneous value of the function $\mu_{\text{ramp}}^{\text{QLV}}(t,\gamma_{o,0})$, defined as $\mu_{\text{ramp}}^{\text{QLV}}(t^*,\gamma_{o,0})$.
The point $\mu_{\text{ramp}}^{\text{QLV}}(t^*,\gamma_{o,0})$ is the maximum of the relaxation curve. In the linear model of \Eq{eq:pronymu}, the maximum of the relaxation curve (obtained for a step-strain test) is equal to the instantaneous shear modulus $\mu_0$ in \Eq{eq:mu0}. For a ramp test, the instantaneous response is related to the instantaneous modulus through the formula in \Eq{eq:k0ramp}, where the elastic constants $k_i$ are replaced by $\mu_i$ for $i=\{1,\dots,n\}$, respectively.

We can now investigate how the non-linear terms in \Eq{eq:rf-qlv} affect the instantaneous response of the torque. To quantify this effect, we compute the following function:
\begin{equation}\label{eq:modifiedQLVmu0}
	\begin{split}
		\tilde{\mu}_{0\text{ramp}}^{\text{QLV}}&=\dfrac{\mu_{\text{ramp}}^{\text{QLV}}(t^*,\gamma_{o,0})-\mu_{\infty}}{\mu_0}\\
		&=\nu_1^{-1}\left( 1-e^{-\nu_1}\right) +	\frac{2}{9}  (1+\nicefrac{2 c_2}{\mu_0}) \nu_1^{-3} e^{-\nu_1 } \left(e^{\nu_1 } \left(\nu_1 ^2-4 \nu_1 +6 \right)-2 (\nu_1 +3) \right) \gamma_{o,0}^2
	\end{split}
\end{equation}
where $\nu_1=\nicefrac{t^*}{\tau_1}$.

\begin{figure}[!b]
	\centering
	\includegraphics[width=0.7\linewidth]{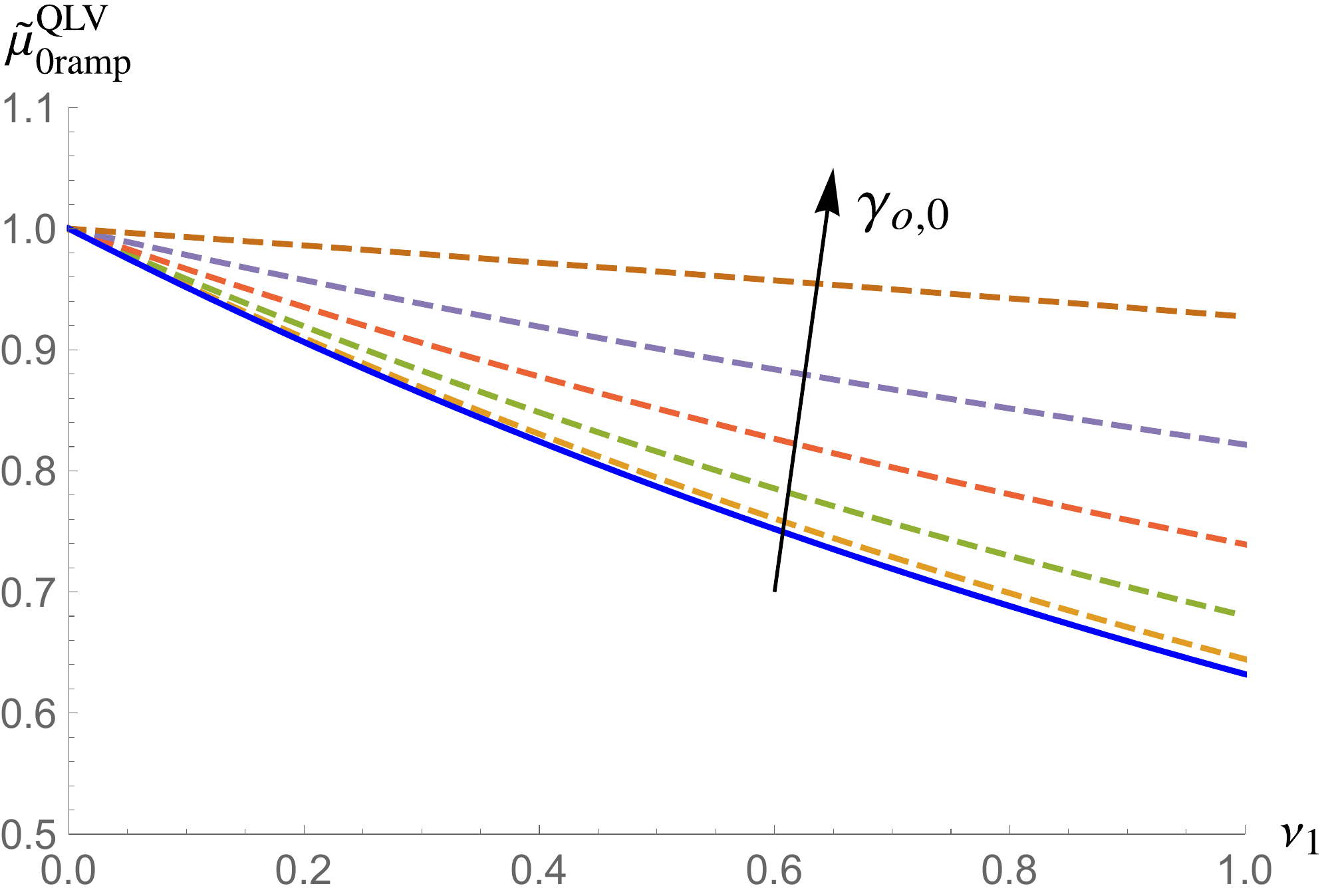}
	\caption{Effect of the rising time $t^*$ and of the strain level $\gamma_{o,0}$ on the maximum of the relaxation curve of the torque. The function in \Eq{eq:modifiedQLVmu0} is plotted with respect to $\nu_1=t^*/\tau_1$ for different values of strain $\gamma_{o,0}=\{0.02,\pi/5,\pi/4,\pi/3,\pi/2,\pi\}$. The solid blue line is obtained from the linear model, see \Eq{eq:k0ramp} and does not depend on the level of strain.}
	\label{fig:mu0-qlv}
\end{figure}

In \Fig{fig:mu0-qlv}, we plot \Eq{eq:modifiedQLVmu0} with respect to $\nu_1=\nicefrac{t^*}{\tau_1}$ and for different values of $\gamma_{o,0}$. The parameter $\nu_1$ spans from 0 to 1, where the value $0$ corresponds to perfect step input ($t^*=0$) and the value $1$ corresponds to a ramp input when the rising time of the ramp is equal to the characteristic relaxation time of the tissue ($t^*=\tau_1$).

The curves in \Fig{fig:mu0-qlv} show that the higher $\gamma_{o,0}$, the higher is the peak of the relaxation curve. For $\gamma_{o,0}=0.02$ (solid blue line), Eq. \eqref{eq:modifiedQLVmu0} recovers the relaxation curve in \Eq{eq:maxTorLin}. In the small deformation regime $\gamma_{o,0}\ll 1$, the non-linear terms in \Eq{eq:modifiedQLVmu0} are very small and as expected the predictions of the QLV model recover those of the linear model. \\
When $t^*\rightarrow 0$ (\ie $\nu_1\rightarrow 0$), the ramp is infinitely fast and is very close to be a step. In this limit \Eq{eq:rf-qlv} reduces to \Eq{eq:pronymu} and the linear and quasi-linear viscoelastic theories predict the same results for any value of $\gamma_{o,0}$. This limit corresponds to the point $(0,1)$ in \Fig{fig:mu0-qlv}.

The effect of the rising time $t^*$ (and therefore of the strain-rate if $\gamma_{o,0}$ is fixed) on the relaxation curve is accounted by the parameter $\nu_1$: the slower the test, the lower the peak of the experimental relaxation curve in \Eq{eq:rf-qlv} and the bigger is the difference between the predictions of the linear and the QLV model.\\
When $t^*\rightarrow\infty$, \ie for very slow (quasi-static) tests, $\tilde{\mu}_{0\text{ramp}}^{\text{QLV}}\rightarrow0$, which means that only the quasi-static elastic properties of the material, \ie $\mu_{\infty}$, can be estimated. This would correspond to a quasi-static experiment.

Moreover, from \Fig{fig:mu0-qlv} we can quantify the influence of the quadratic term in \Eq{eq:modifiedQLVmu0}. For small values of $\nu_1$, the quadratic contribution is minimal even for large values of strain. Its effect increases with both $\nu_1$ and $\gamma_{o,0}$, as expected.

Note that $\tilde{\mu}_{0\text{ramp}}^{\text{QLV}}$ decreases as $\nu_1$ increases, but increases with $\gamma_{o,0}$. These two opposite effects can lead to an apparent compensation, \ie constant value of $\tilde{\mu}_{0\text{ramp}}^{\text{QLV}}$, when slow experiments are performed in the large deformation range.

Finally, we note that the second Mooney-Rivlin parameter $c_2$ enters \Eq{eq:rf-qlv} through the quadratic term $\gamma_{o,0}^2$ of the function $\mu_{\text{ramp}}^{\text{QLV}}(t,\gamma_{o,0})$.
As we highlighted above, the quadratic term vanishes in both the limits $t^*\rightarrow0$ (step-strain test) and $t\rightarrow\infty$ (elastic equilibrium). 
Therefore, it is difficult to determine the parameter $c_2$ from the torque data. The identification of $c_2$ requires information on the normal force.\\
In conclusion, the relaxation curve of the torque allows us to estimate the instantaneous and the long-term moduli $\mu_0$ and $\mu_{\infty}$, respectively, the moduli $\mu_i$ and the relaxation times $\tau_i$ from a step or a ramp test.


\subsubsection{Normal Force}\label{sec:N}
In this section we derive the analytical expression of the relaxation curve for the normal force predicted by the QLV model. As in the previous section, we consider both the ramp test and the step-strain test scenarios. Furthermore, we will show how to use the normal force data measured from the two tests to estimate the second Mooney-Rivlin parameter $c_2$. \\
We recall that the only non-zero component of the infinitesimal strain tensor is the shear component $\varepsilon_{\theta z}$. It follows that all three normal components of the Cauchy stress tensor are zero, particularly $\sigma_{zz}=0,\forall t$. Therefore, in the small deformation regime the linear theory predicts a zero normal force response for any strain history input. \\
On the other hand, in the large deformation regime the QLV theory predicts a non-zero normal force for an incompressible Mooney-Rivlin viscoelastic material under torsion, see \Eq{eq:qlv-force-gamma}.
By substituting the deformation history \eqref{eq:ramp2} into \Eq{eq:qlv-force-gamma}, we calculate the relaxation curve for the normal force in response to a ramp input. Upon integrating, we obtain:
\begin{equation}\label{eq:qlv-Fz}
	\begin{split}
		N(t,\gamma_{o,0})=&-\frac{\pi}{4}  r_o^2\left( \left(1+2 c_2/\mu_0\right) \mu _{\infty }
		+2 \sum_{i=1}^{n} \mu_i \nu_i^{-2} e^{-\frac{t}{\tau _i}} \left(2 c_2/\mu_0 \left(e^{\nu _i} \left(\nu _i-1\right)+1\right)+e^{\nu _i}-\nu _i-1\right) \right) \gamma _{o,0}^2\\
		&-\frac{\pi}{9} \left(1+2 c_2/\mu_0\right) r_o^2 \sum_{i=1}^{n} \mu _i \nu_i^{-4}  e^{-\frac{t}{\tau _i}} \left(e^{\nu _i} \left(\nu _i^2-6 \nu _i+12\right)-\nu _i^2-6 \nu _i -12\right) \gamma _{o,0}^4,
	\end{split}
\end{equation}
which is valid for $t>t^*$ and where $\nu_i=t^*/\tau_i$.
From \Eq{eq:qlv-Fz} we observe that the normal force $N$ is given the sum of two terms proportional to $\gamma _{o,0}^2$ and $\gamma _{o,0}^4$, respectively.
Therefore, in the small deformation limit ($\gamma_{o,0}\!\rightarrow0$) \Eq{eq:qlv-Fz} recovers the predictions of the linear theory, \ie $N\!=\!0, \forall t$.
According to the definition in \eqref{eq:prony}, the relaxation function associated to the normal force is $\sigma_{zz}(t,\gamma_{o,0})/\varepsilon_0$, where $\sigma_{zz}(t,\gamma_{o,0})=-N(t,\gamma_{o,0})/(\pi r_o^2)$ and $\varepsilon_0=\gamma_{o,0}/2$, namely:
\begin{equation}\label{eq:muN}
	\mu_N(t,\gamma_{o,0})=-\frac{2 N(t,\gamma_{o,0})}{\pi r_o^2 \gamma_{o,0}}.
\end{equation}
In the limit $t^*\rightarrow0$ (step strain input), \Eq{eq:muN} reduces to:
\begin{equation}\label{eq:muN-step}
	\lim_{t^*\rightarrow 0}\mu_N(t,\gamma_{o,0})=\frac{\gamma_{o,0}}{2}\,\left( 1+2\frac{c_2}{\mu_0}\right) \mu(t) , \quad \text{with} \quad \mu(t)=\mu_{\infty}+\sum_{i=1}^n \mu_i e^{-\frac{t}{\tau_i}}.
\end{equation}
The factor $\gamma_{o,0}/2$ represents the dependence on the deformation, while the factor \mbox{$(1+2c_2/\mu_0)\mu(t)$} represents the dependence on time, which is indeed strain-independent. Thus, it follows that different relaxation curves obtained for different values of the strain $\gamma_{o,0}$ do not overlap.
The functional dependence of $\mu_N$ on the strain $\gamma_{o,0}$ is dictated by the form of the elastic constitutive model. In this particular case, the choice of a Mooney-Rivlin model yields a linear dependence on $\gamma_{o,0}$ in \Eq{eq:muN-step}.
However, the relaxation curves display the same exponential decay in time, which is dictated by the choice of the rheological model. For a generalised Maxwell model, the decay is exponential according to the Prony series in \Eq{eq:muN-step}.
If the experimental curves do not display the same decay in time, then the QLV model will not fit the data accurately and more advanced/non-linear models should be considered \cite{christensen2012theory}.

Since $N(t)$ depends on the deformation through quadratic terms in $\gamma _{o,0}$ and the aim here is to illustrate how the relaxation curve of the normal force is affected by the level of deformation and the ramp phase, we choose to introduce the following function:
\begin{equation}\label{eq:fN}
	\begin{split}
		f_N(t,\gamma_{o,0})=&-\frac{2 N(t)}{\pi r_o^2 \gamma_{o,0}^2}\\
		=&\frac{1+2 c_2/\mu_0}{2} \mu _{\infty }
		+\sum_{i=1}^{n} \mu_i \nu_i^{-2} e^{-\frac{t}{\tau _i}} \left(2 c_2/\mu_0 \left(e^{\nu _i} \left(\nu _i-1\right)+1\right)+e^{\nu _i}-\nu _i-1\right)\\
		&+\frac{1+2 c_2/\mu_0}{9} \sum_{i=1}^{n} \mu _i \nu_i^{-4}  e^{-\frac{t}{\tau _i}} \left(e^{\nu _i} \left(\nu _i^2-6 \nu _i+12\right)-\nu _i^2-6 \nu _i -12\right) \gamma _{o,0}^2,
	\end{split}
\end{equation}
which is valid for $t>t^*$. By taking the limit $t\rightarrow\infty$, \Eq{eq:fN} reduces to:
\begin{equation}\label{eq:fN-inf}
	f_{N_{\infty}}=\lim_{t\rightarrow\infty} f_N(t,\gamma_{o,0})= \left(1/2+c_2/\mu_0\right) \mu _{\infty }.
\end{equation}
The limiting value in \Eq{eq:fN-inf} explicitly depends on the second Mooney-Rivlin coefficient $c_2$. Therefore, experimentally we can determine $c_2$ by performing a ramp (or a step) test and by measuring the asymptotic value of the normal force curve as $t\rightarrow\infty$.
We point out that the parameters $c_2$ appears explicitly in the limiting value (as $t\rightarrow\infty$) of the normal force only (it does not appear in the asymptotic value on the torque). This is a consequence of the fact that the vertical force is associated to the change in area of the section of the cylinder. Furthermore, this is consistent with the fact that $c_2$ is the parameter associated to the second invariant $I_2$ of the strain tensor $\textbf{B}$, which indeed accounts for the changes in the area of the material due to the deformation.\\
Finally, to show the influence of the rising time $t^*$ and of the level of strain at the end of the ramp $\gamma_{o,0}$ on the function $f_N$, we plot $f_N/\mu_0$ for different values of $\nu_1$ and $\gamma_{o,0}$ in \Fig[a]{fig:fnnu-gamma} and \Fig[b]{fig:fnnu-gamma}, respectively.

\begin{figure}[htb]
	\subfigure[]{\includegraphics[width=0.49\linewidth]{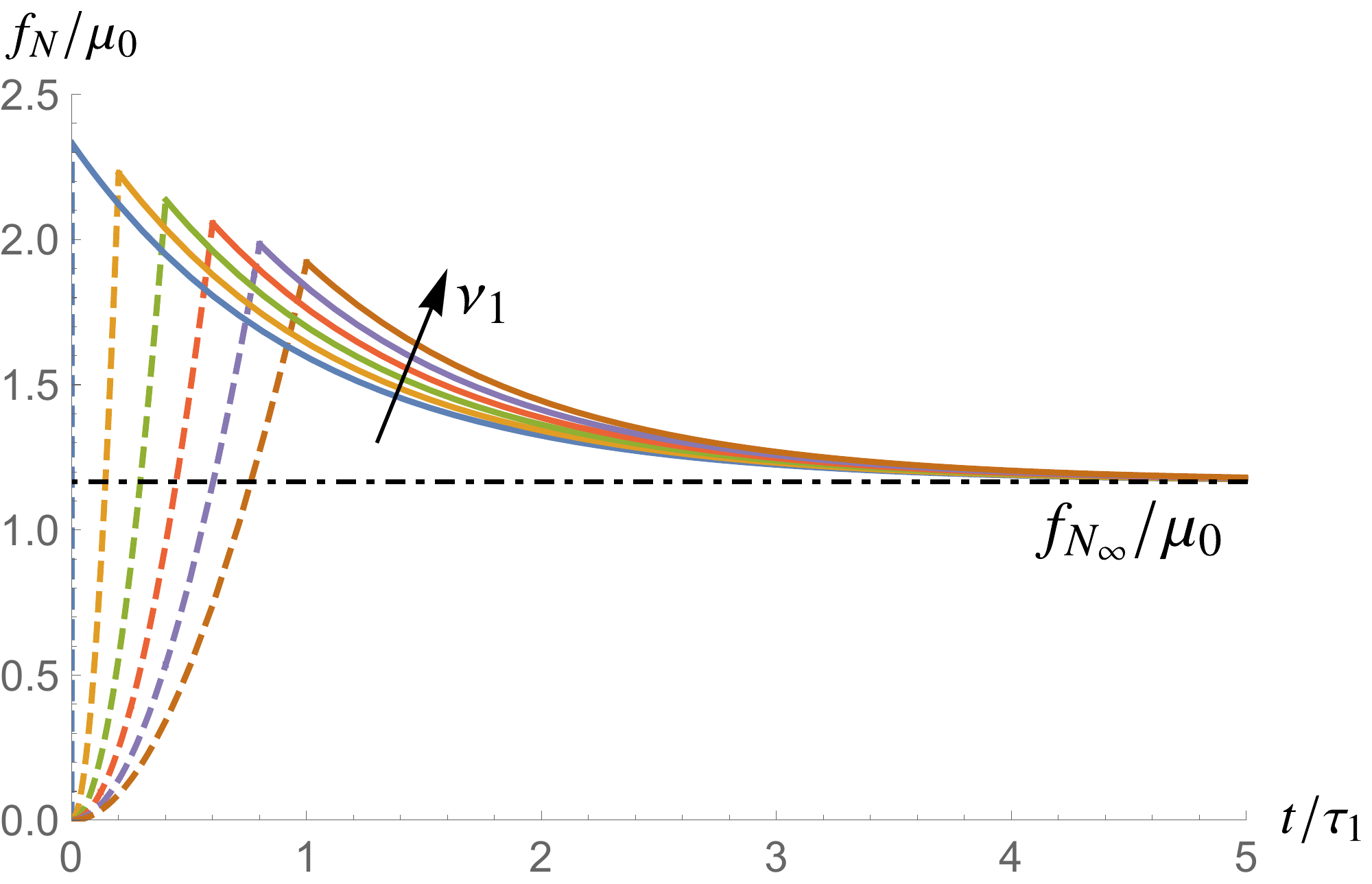}}\hfill
	\subfigure[]{\includegraphics[width=0.49\linewidth]{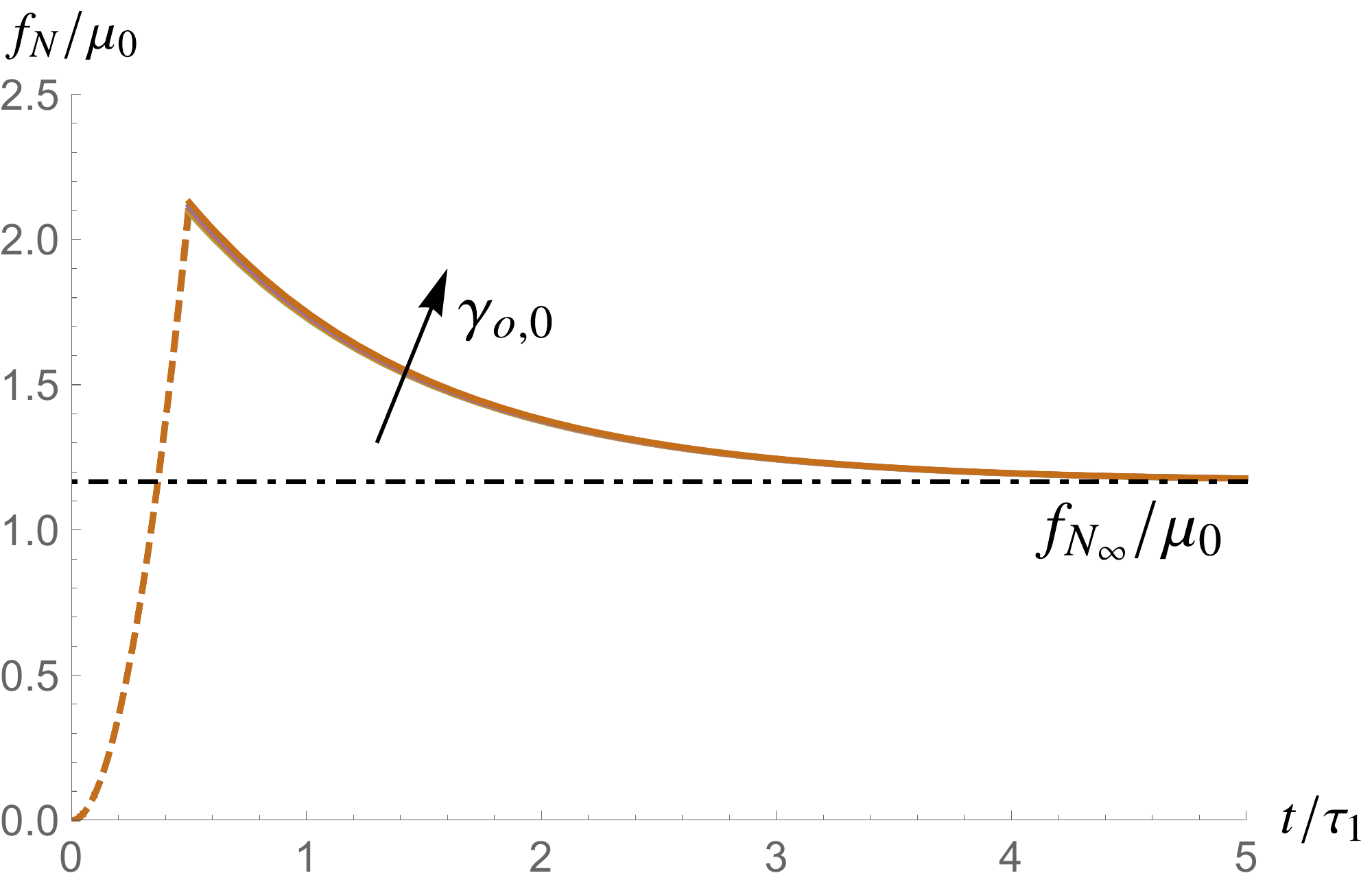}}
	\caption{Effect of the rising time (a) and of the strain level (b) on the relaxation curve of the normal force in \Eq{eq:fN}. The following parameters are fixed: $n=1$, $\tau_1=1$, $\mu_1=\mu_{\infty}=1$, $c_2/\mu_0=2/3$.
		In (a) we fix $\gamma_{o,0}=0.02$ and let $\nu_1$ spanning over $\{0,0.2,0.4,0.6,0.8,1\}$. $\nu_1=0$ represents a perfect step test and $\nu_1=1$ represents a ramp test with rising time $t^*=\tau_1$. In (b) we fix $\nu_1=1/2$ and let $\gamma_{o,0}=\{0.02,\pi/5,\pi/4,\pi/3,\pi/2,\pi\}$.}
	\label{fig:fnnu-gamma}
\end{figure}

\Fig[b]{fig:fnnu-gamma} shows that the final level of strain reached at the end of the ramp has a negligible effect on the relaxation curves of the normal force. However, for a fixed strain level, the slower is the ramp, the lower is the peak of the relaxation curve, similarly as we observed for the relaxation curves of the torque (\Fig{fig:muqlvtnu}).

In conclusion to fully characterise the viscoelastic behaviour of a soft tissue  in torsion (that obeys a Mooney-Rivlin hyperelastic law), according to the QLV theory, we only need to perform a single step-strain test. This test can be performed by using a rheometer that gives access to two sets of data, the torque and the normal force.  Moreover, as we showed in \Fig{fig:mu0-qlv} and \Fig[b]{fig:fnnu-gamma}, the level of strain does not affect the relaxation curves of the torque and the normal force for a step test. Therefore, if the rheometer can achieve high strain rates and if we are only interested in estimating $\mu_0,\mu_{\infty},\mu_i$ and $\tau_i$ (for instance if the material is Neo-Hookean, \ie $c_2=0$), the test can be performed in the small deformation regime and the parameters can be estimated by fitting the torque data with \Eq{eq:pronymu}.
Otherwise, the test should be modelled as a ramp-test in the large deformation regime (see \Sec{sec:largedef}). In this case, from the relaxation curve of the torque in \Eq{eq:rf-qlv} we can estimate the long-term shear modulus $\mu_{\infty}$ from the value of the data as $t\rightarrow\infty$, according to \Eq{eq:qlv-inf}. We can obtain all the other moduli $\mu_i$ and the relaxation times $\tau_i$ by fitting the relaxation curve in \Eq{eq:rf-qlv}. Finally, we can estimate the parameter $c_2$ from the value of the normal force data as $t\rightarrow\infty$, according to \Eq{eq:fN-inf}.

\section{Conclusions}
In this Chapter we reviewed the foundations of linear viscoelasticity and the theory of Quasi Linear Viscoelasticity (QLV). With the aim of providing a fitting procedure for the QLV model and estimating the viscoelastic properties of a soft tissue, we considered the torsion of a soft solid cylinder and we wrote the governing equations of the viscoelastic problem for a tissue that elastically behaves as a Mooney-Rivlin material, such as the brain. We derived the analytical predictions of the relaxation curves for the torque and the normal force necessary to twist a cylindrical sample. We considered two experimental scenarios: the step test, where the tissue is instantaneously deformed and held in position, and the ramp test, where the tissue is deformed in a finite time and then held in position. These tests are commonly performed to characterise the time-dependent properties of soft tissues and allow to investigate their stress relaxation behaviour. We investigated the effect of the strain level and rising time of the ramp on the relaxation curves of the torque and the normal force. Our results show that in a step test, the linear and the QLV models predict the same relaxation curves for the torque. However, when the strain input is in the form of a ramp function, the non-linear terms appearing in the QLV model affect the relaxation curve of the torque depending on the strain level attained at the end of the loading phase (see \Fig{fig:mu0-qlv}). In particular, the higher is the strain level, the higher is the maximum of the relaxation curve, whilst the equilibrium value remains constant and unchanged. \\
The linear model predicts a zero normal force $\forall t$, whilst the QLV model predicts a non-zero normal force that depends on $\gamma_{o,0}^2$ and $\gamma_{o,0}^4$. Our results show that the relaxation curve of the normal force depends on the level of strain both for a step test (see \Eq{eq:muN-step}) and a ramp test (see \Eq{eq:muN}). Although the contributions of the non-linear terms are negligible for a Mooney-Rivlin material (see \Fig[b]{fig:fnnu-gamma}), their effect might be relevant for materials that obey a different elastic law (\eg Fung, Ogden, Gent, etc.).\\
Finally, our results provide useful guidelines to accurately fit QLV models in view of estimating the viscoelastic properties of soft tissues. We showed how to use the data from a ramp test in torsion (\ie the relaxation curves of the torque and the normal force) to estimate the constitutive parameters of a Mooney-Rivlin viscoelastic tissue. \\

\end{document}